\documentclass[aps, prl, reprint, groupeaddress, floatfix, longbibliography]{revtex4-1}
\usepackage[T1]{fontenc}
\usepackage{amsmath}
\usepackage{amssymb}
\usepackage{epsfig}
\usepackage{graphicx}
\usepackage{braket}
\usepackage[usenames,dvipsnames]{xcolor}
\usepackage{bbold}
\usepackage[normalem]{ulem}

\newcommand*\adag{\ensuremath{\hat{a}^\dagger}}

\newcommand*\bdag{\ensuremath{\hat{b}^\dagger}}

\newcommand*\Adag{\ensuremath{\hat{A}^\dagger}}
\newcommand*\Aop{\ensuremath{\hat{A}}}

\newcommand*\os{\ensuremath{\omega_\mathrm{s}}}
\newcommand*\oi{\ensuremath{\omega_\mathrm{i}}}

\newcommand*\fs{\ensuremath{f^{(\mathrm{s})}}}
\newcommand*\fid{\ensuremath{f^{(\mathrm{i})}}}
\newcommand*\fc{\ensuremath{f^{(\mathrm{c})}}}

\newcommand*\gauss{\includegraphics[height=6pt]{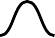}}
\newcommand*\hermite{\includegraphics[height=6pt]{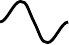}}

\definecolor{myblue}{rgb}{0.0, 0.25, 0.75}
\definecolor{mygreen}{RGB}{0, 115, 77}

\begin{document}

\title{Photon temporal modes: a complete framework for quantum information science}

\author{B. Brecht$^{1\star}$, Dileep V. Reddy$^2$, C. Silberhorn$^1$, and M. G. Raymer$^2$.}

\affiliation{$^1$Integrated Quantum Optics, Applied Physics, University of Paderborn, Warburger Strasse 100 33098, Paderborn, Germany}	

\affiliation{$^2$Oregon Center for Optics, Department of Physics, University of Oregon, Eugene, Oregon 97403, USA}

\date{\today}

\begin{abstract}
Field-orthogonal temporal modes of photonic quantum states provide a new framework for quantum information science (QIS). They intrinsically span a high-dimensional Hilbert space and lend themselves to integration into existing single-mode fiber communication networks. We show that the three main requirements to construct a valid framework for QIS -- the controlled generation of resource states, the targeted and highly efficient manipulation of temporal modes and their efficient detection -- can be fulfilled with current technology. We suggest implementations of diverse QIS applications based on this complete set of building blocks. %
\end{abstract}


\maketitle

\section{Introduction}
Quantum information science (QIS) offers means for storing, transmitting and processing information in ways not achievable using classical information technology. Examples of the benefits of QIS are unconditionally secure communication, ultra-precise metrology beyond classical limits, and superior computational algorithms. %

While all of those can theoretically be realized using only photons, it is generally accepted that quantum computation will be implemented in material systems, whereas quantum communication and information transfer across a distributed quantum network -- a so-called ``quantum internet'' \cite{Kimble:2008if} -- will be based on photons. Strongly interacting material systems, which can be controlled with outstanding precision, facilitate the implementation of stationary logical processors and quantum memories. The latter are an indispensable building block for long-distance entanglement distribution via quantum repeaters, which in turn is inextricably linked with secure long-distance quantum communication. Photons, in contrast, interact only weakly with themselves and their environment, meaning that they experience very low decoherence. Thus, they are naturally suited for carrying fragile quantum information over transmission lines in a network. The remaining challenge for these hybrid network architectures is the efficient interfacing of flying qubits (photons) and stationery qubits (material systems), which is complicated by the fact that most practical material systems have stringent requirements on the photon spectral-temporal amplitude. Thus, small photonic co-processor units that facilitate, for instance, the coherent re-shaping of photons in time and frequency must be available. Note that these do not necessarily have to fulfill the more stringent demands of fault-tolerant quantum computation to be practical and therefore, as we show, can be realized with current technology. %

In this paper we introduce a practical framework for photonic quantum information science. Our framework exploits temporal modes (TMs) of single photon states -- field-orthogonal broadband wave-packet states -- that have to date not been demonstrated to enable a viable basis for quantum information encoding. In particular, we complement existing knowledge with all missing building blocks, which are needed to demonstrate that TMs satisfy the three major requirements for the implementation of the photonic subsystems of large-scale quantum networks: firstly, for the preparation of good signal carriers, appropriate resource states have to be generated and completely characterized with high reliability and flexibility; secondly, the subsequent processing of quantum information in co-processor units requires that controlled operations can be implemented; finally, efficient detection schemes, which enable faithful information readout, must be available. %

We expect that the TM framework for photonic quantum information will open avenues towards the realisation of practical QIS applications. One such application is the boson sampler \cite{Aaronson:2011ho,Spring:2013do,Broome:2013bv,Crespi:2013fu,Tillmann:2013jv}, which, though not on par with the requirements for fault-tolerant quantum computation, may soon show performance beyond the capabilities of state-of-the-art classical computer, which are pushed to their limits by linear optical networks with about 100 modes, of which only 30 are occupied. Our new TM paradigm may offer improved methods to construct large networks with reduced switching losses, which are currently thought to be the main limiting factor when considering the scalability of photonic quantum processing \cite{Li:2015ue}. %

In the following we first introduce the basic concepts of our framework by formally defining TMs and their use as an information-encoding basis. Then we briefly review the current state-of-the-art of generating TMs with ultrafast parametric down-conversion, where we will outline why existing sources do not yet fulfill the requirements for QIS with TMs. After this, we highlight recent developments in TM manipulation, which serve as starting point for the definition of the complete TM framework. The key enabling findings for this are our recent results, which introduce means for sorting TMs with high efficiency and selectivity in excess of 99.5 percent. This high efficiency of the ``quantum pulse gate'' operation can be achieved by dispersion-engineered, multi-stage frequency conversion driven by spectrally-temporally shaped laser control pulses. We then present new concepts and components, which enable the establishment of the complete TM framework. In particular, we design the flexible generation of entangled resource states of arbitrary, user-defined dimension, we introduce TM quantum-state tomography of single-photon as well as photon-pair states to verify the successful state generation, and we establish concrete applications for QIS. We show that all operations necessary to implement photonic co-processors and quantum communication applications can be implemented with TMs. We conclude the paper with a discussion of the experimental challenges and limitations of our framework. %

\section{Fundamental Concepts}
Starting from a very general point-of-view, we note that light has four degrees of freedom (DOF), any of which could be used to encode quantum information: these are the helicity and the three components of the momentum vector. In a beam-like geometry these may be stated as polarization, transverse mode profile (encompassing two DOFs), and energy (that is, frequency). From these DOFs, polarization is most widely applied in quantum information processing. The generation of polarization-entangled Bell states \cite{Kwiat:1995ck} as resource states is nowadays an established experimental method. Two orthogonal polarization modes can easily be separated by means of using polarizing beamsplitters, and proper gate operations are readily implemented with linear optical elements such as waveplates, (polarizing) beamsplitters and detectors. However, polarization intrinsically spans a mere two-dimensional Hilbert space, and thus cannot exploit the true potential of QIS, which, in certain cases such as quantum key distribution, benefits from higher-dimensional Hilbert spaces \cite{BechmannPasquinucci:2000ug,Cerf:2002fp}. %

The second DOF, transverse mode profile, has received considerable attention recently, as it has become apparent that the orbital-angular-momentum (OAM) states of light are a useful basis for encoding information \cite{Mair:2001fd,Leach:2010tt,Dada:2011do} and can be efficiently sorted with time-stationary linear optical elements \cite{Berkhout:2010cb}. They have been used recently to demonstrate, for instance, enhanced security and bitrate in quantum communication \cite{Groblacher:2006ec,Barreiro:2008jl,Leach:2012gu}. Still, the OAM basis has three drawbacks limiting its current value for some QIS applications: first, it is inherently incompatible with existing single-mode fiber networks because information is encoded onto different spatial field distributions; second, it is susceptible to medium perturbations such as turbulence, which affects free-space links; and third, the generation of OAM states with a tailored structure, for instance a well-defined number of modes, is as of yet an unsolved problem. %

Only recently has the final DOF of light -- energy, that is frequency -- been recognized as an underutilized resource for QIS. Because frequency and time are conjugate variables, we call a set of overlapping but orthogonal broadband wave-packet modes by the name ``temporal modes'' (TMs). In a coherent-beam-like or single-transverse-mode guided wave geometry, TMs form a complete basis for representing an arbitrary state in the energy degree of freedom \cite{Titulaer:1966vb}. TMs overlap in time and frequency, yet are field-orthogonal. In this respect, they are analogous to transverse spatial modes, yet they possess distinct advantages. Since all TMs ``live'' inside the same spatial field distribution, they are naturally suited for use with highly efficient and experimentally robust waveguide devices and existing single-mode fiber networks. In addition, they are insensitive to stationary or slowly-varying medium perturbations such as linear dispersion, due to their overlapping spectra, making them suitable for real-world applications. %

While the TM concept applies to any states of light (e.g. squeezed quadrature states \cite{Roslund:2013cb,Averchenko:2014dn}), we restrict ourselves to single-photon states to keep this paper concise and readable. In this context, TMs are a complete mode set for expanding the electromagnetic field and, in addition, can be regarded as a complete set of quantum states for single photons. %

\subsection{Temporal modes for single-photon states}
\begin{figure}
    \centering
    \includegraphics[width=\linewidth]{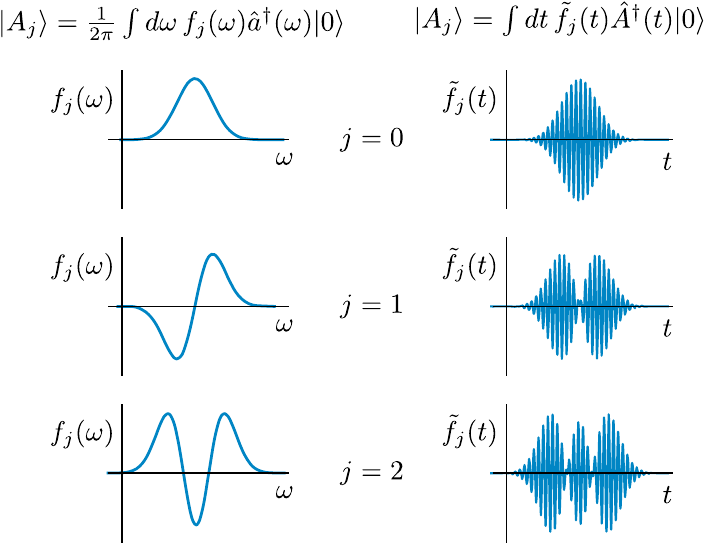}
    \caption{First three members of a TM basis in the frequency domain (left) and the time domain (right).}
    \label{fig:temporal_modes}
\end{figure}

For a fixed polarization and transverse field distribution (e.g. in a beam-like geometry), a single-photon quantum state in a specific TM can be expressed as a coherent superposition of a continuum of single-photon states in different monochromatic modes: 
\begin{equation}
    \ket{A_j} = \int \frac{d\omega}{2\pi}\, f_j(\omega)\adag(\omega)\ket{0}.
\end{equation}
Here, $\adag(\omega)$ is the standard monochromatic creation operator and $f_j(\omega)$ is the complex spectral amplitude of the wave packet. By Fourier transform, this same state can be expressed as a coherent superposition over many possible ``creation times'', and then reads
\begin{equation}
    \ket{A_j} = \int dt\, \tilde{f}_j(t)\Adag(t)\ket{0}\equiv\Adag_j\ket{0},
    \label{eq:temporal_state}
\end{equation}
where we used the definition
\begin{equation}
    \adag(\omega)=\int dt\, e^{\imath\omega t}\Adag(t);\:
    \Adag(t) = \int \frac{d\omega}{2\pi}\, e^{-\imath\omega t}\adag(\omega).
\end{equation}
In Eq. (\ref{eq:temporal_state}), $\tilde{f}_j(t)$ is the temporal shape of the wave packet (defined as the Fourier transform of $f_j(\omega)$) and $\Adag(t)$ creates a photon at time $t$. We also defined a so-called broadband-mode operator
\begin{equation}
    \Adag_j=\int dt\,\tilde{f}_j(t)\Adag(t)=\frac{1}{2\pi}
    \int d\omega\,f_j(\omega)\adag(\omega),
    \label{eq:broadband_operator}
\end{equation}
which creates the wave-packet state $\ket{A_j}$. In Fig. \ref{fig:temporal_modes}, we exemplarily plot the first three members of a TM basis, chosen for illustration to be a family of Hermite-Gaussian functions of frequency. With this, it is possible to express every single-photon temporal wave-packet quantum state $\ket{\Psi}$ in a basis of TMs as a superposition of wave-packet states,
\begin{equation}
    \ket{\Psi}=\sum_{j=0}^\infty c_j\hat{A}^\dagger_j\ket{0},
\end{equation}
with complex-valued expansion coefficients $c_j$. 

We want to highlight that, although they fully overlap in polarization, space, frequency and time, TMs are orthogonal with respect to a frequency (time) integral
\begin{equation}
    \frac{1}{2\pi}\int d\omega\,f_j^*(\omega)f_k(\omega) = 
    \int dt\,\tilde{f}_j^*(t)\tilde{f}_k(t) = \delta_{jk}.
\end{equation}
They also obey bosonic commutation relations \cite{Titulaer:1966vb,Smith:2007dq}
\begin{equation}
    [\Aop_i,\Adag_j] = \delta_{ij}
\end{equation}
just as do the well-known monochromatic creation operators. %

\subsection{Quantum information encoding with TMs}
Deploying TMs for quantum information encoding is an appealing prospect, because TMs span an infinite dimensional Hilbert space. This has been shown to facilitate increased information capacity per photon and increased security in quantum communication \cite{Groblacher:2006ec,Barreiro:2008jl,Leach:2012gu} when compared to two-dimensional encoding. The carriers of information in a $d$-dimensional Hilbert space are typically called ``qudits''.

We define a TM qudit as a coherent superposition of $d$ TM
states:
\begin{equation}
  \ket{\psi}_\mathrm{TM}^{d}=\sum_{j=0}^{d-1}\alpha_j\ket{A_j}.
  \label{eq:TM_qudit}
\end{equation}

\begin{figure}
    \centering
    \includegraphics[width=\linewidth]{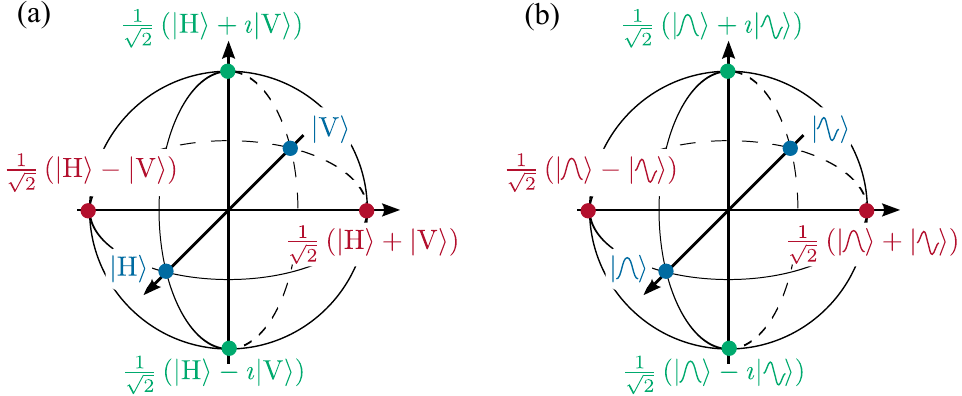}
    \caption{(a) Poincar\'e sphere. The logical ``0'' and ``1'' of a
      polarization qubit can be encoded in any two diametrically
      opposite points on the sphere. Typically, horizontal and
      vertical polarization are deployed. (b) Bloch sphere for TM
      qubits. Any two orthogonal TMs and their coherent superpositions
      may be used to encode TM qubits. In this example, the TMs are a
      zeroth and first order Hermite-Gaussian.}
    \label{fig:bloch_spheres}
\end{figure}

To highlight the formal similarity of TMs with other encoding bases, we start by discussing TM qubits. The most common implementation of a photonic qubit is the polarization qubit, which can be written as $\ket{\psi}=\alpha\ket{\mathrm{H}}+\beta\ket{\mathrm{V}}$. Here, $\ket{\mathrm{H}}$ and $\ket{\mathrm{V}}$ denote horizontal and vertical polarization, respectively, and $|\alpha|^2+|\beta|^2=1$. Commonly, a polarization qubit is represented as a point on the surface of a Poincar\'e sphere as sketched in Fig. \ref{fig:bloch_spheres}(a).

In analogy to this, the definition of a TM qubit requires two orthogonal states with which we associate the logical ``0'' and ``1''. Without loss of generality, we can consider zeroth-order and first-order Hermite-Gaussian  functions of frequency to define the TMs, labeled $\gauss$ and $\hermite$, and consequently write

\begin{equation}
    \ket{0}\equiv\ket{\gauss},\hspace{0.5cm}\ket{1}\equiv\ket{\hermite}.
\end{equation}

Then, a TM qubit is given by
\begin{equation}
    \ket\psi_\mathrm{TM}\equiv\alpha\ket\gauss + \beta\ket\hermite,
    \label{eq:tm_qubit}
\end{equation}
where again $|\alpha|^2+|\beta|^2=1$. Similar to polarization qubits, the TM qubit is best visualized as a point on the surface of a Bloch sphere as shown in Fig. \ref{fig:bloch_spheres}(b). 

\subsection{Mutually unbiased bases}
Sets of bases, for which the overlap between a basis vector of one basis with any basis vector from any of the other bases has the same absolute value, are called mutually unbiased bases (MUBs) \cite{Schwinger:1960wq}. They lie at the heart of QIS applications such as quantum key distribution \cite{Bennett:1984wv} or quantum state tomography \cite{Smithey:1993er}. The physical meaning of MUBs is the following: if a certain quantum state is an eigenstate of one basis then a measurement in any other MUB yields a uniformly random result yielding no information. Using polarization states, the three sets of Stokes vectors denoting horizontal and vertical, diagonal and anti-diagonal as well as left- and right-circular light form the typically used MUBs.

\begin{figure}
    \centering
    \includegraphics[width=\linewidth]{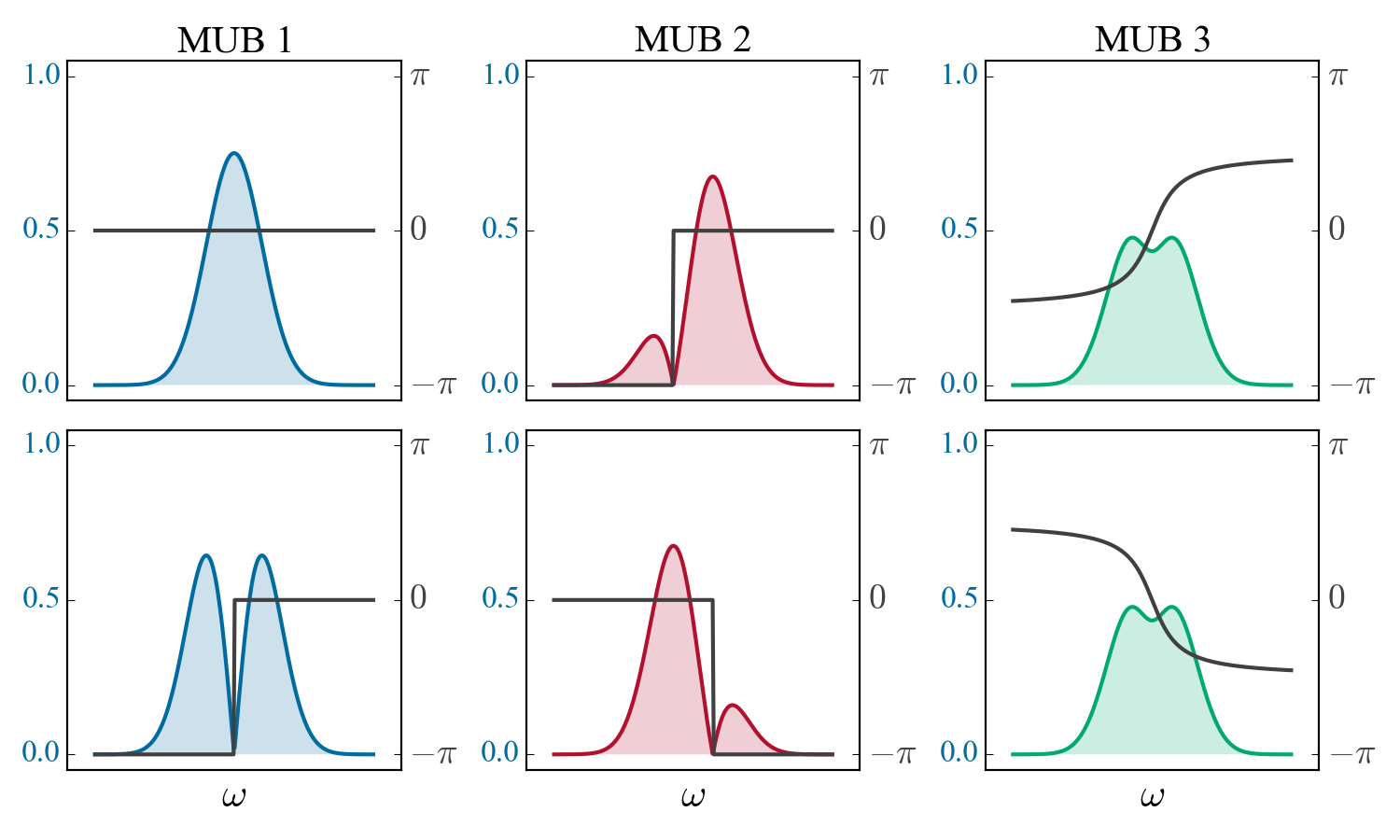}
    \caption{The columns show the three MUBs for a TM qubit, with the
      fundamental TM shapes being a zeroth and first order
      Hermite-Gaussian, respectively. The colored areas are the
      spectral amplitude, whereas the dark lines are the spectral
      phases of the TMs, the color-coding corresponds to
      Fig. \ref{fig:bloch_spheres}(b). Note that in this case, the qubit is
	  encoded in the leftmost basis. }
    \label{fig:mubs}
\end{figure}

For the case of the aforementioned TM qubit from Fig. \ref{fig:bloch_spheres}(b), the basis modes of the three possible MUBs are indicated by the different colors and we explicitly plot them in Fig. \ref{fig:mubs}. The color coding corresponds to Fig. \ref{fig:bloch_spheres}(b). If the qubit was given by $\ket{\psi}_\mathrm{TM}=\ket{\gauss}$, measuring in either the red or green basis results in ``0'' (upper row) or ``1'' (lower row) with a probability of 50\%.

The challenge for TMs is the implementation of a device that facilitates a mode-selective measurement, where the phase coherence plays a particularly important role. For a polarization qubit, an appropriate combination of wave plates and polarizing beamsplitters readily accomplishes the projection onto the respective basis sets. For TMs, the situation is more complicated, since time-stationary operations are not sufficient for mode-selectivity and so-called quantum pulse gates have to be employed \cite{EcksteinA:2011vg,Brecht:2014eg,Reddy:2014bt,Reddy:2015jb}. We return to this point below, where we briefly review the solution to the mode-sorting problem.

\section{State-of-the-art}
In this section, we briefly summarize the current state-of-the-art in generating and manipulating TM states. Typically, the former is realized with parametric down-conversion, whereas the latter can be achieved by deploying TM selective quantum pulse gates.
\subsection{TM structure of photon pair states}
Today, parametric down-conversion (PDC) in optical waveguides is the workhorse for the generation of photon-pair and heralded single-photon states. Notably, PDC generates quantum states with a rich intrinsic TM structure, when ultrafast pulses are deployed as pump \cite{Law:2000wd}. This structure is decoupled from the transverse spatial mode, which is solely determined by the waveguide geometry. It is encoded in the so-called joint spectral amplitude (JSA) of the PDC $f(\os,\oi)$, which can be written as \cite{Grice:1997tk,Keller:1997hj} 

\begin{equation}
    f(\os,\oi) = \alpha(\os,\oi)\cdot\phi(\os,\oi).
\end{equation}
Here, $\alpha(\os,\oi)$ is the pump-envelope function, which encompasses energy conservation and the spectrum of the pump pulses, and $\phi(\os,\oi)$ is the phase-matching function, which describes momentum conservation and depends on the medium dispersion.

\begin{figure*}
    \includegraphics[width=\linewidth]{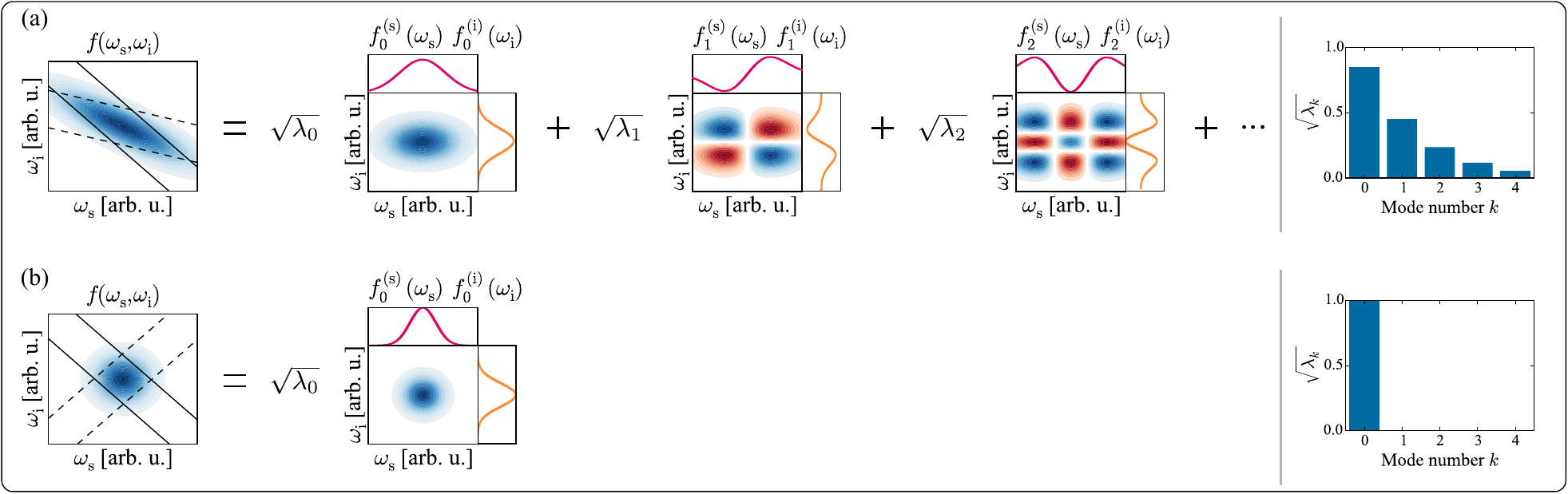}
    \caption{(a) Representation of a general PDC process. The leftmost panel
    shows the JSA $f(\os,\oi)$, which is the product of pump envelope function
    (black solid lines) and the phasematching function (black dashed lines).
    This function is decomposed into two sets of TMs
    $\{\fs(\os)\}$ and $\{\fid(\oi)\}$ with weighting coefficients
    $\sqrt{\lambda_k}$. In the central part, we plot the first three TM pairs. 
    The rightmost panel shows the distribution of expansion coefficients
    $\sqrt{\lambda_k}$. (b) A dispersion-engineered PDC
    process excites only one pair of TMs. The JSA does not exhibit
    any correlations between signal and idler photons. The distribution of
    weighting coefficients $\sqrt{\lambda_k}$ consequently exhibits only a
    single entry greater than zero.}
    \label{fig:pdc_calculation_general}
\end{figure*}

With that, we denote the photon-pair component of the generated state
\begin{equation}
    \ket{\psi}_\mathrm{PDC}=\int d\os d\oi\,f(\os,\oi)
    \adag(\os)\bdag(\oi)\ket{0,0},
    \label{eq:pdc_state}
\end{equation}
where $\adag(\os)$ and $\bdag(\oi)$ are standard monochromatic creation operators for signal and idler photons.

A decomposition of the JSA into two sets of uniquely defined TM basis functions $\{\fs(\os)\}$ and $\{\fid(\oi)\}$, which exhibit pairwise
correlations such that
\begin{equation}
    f(\os,\oi) = \sum_{k=0}^{\infty}\sqrt{\lambda_k}\fs_k(\os)\fid_k(\oi)
    \label{eq:decomposition}
\end{equation}
reveals the underlying TM structure of the PDC state \cite{Law:2000wd}. Here, the expansion coefficients are normalized according to $\sum_k\lambda_k=1$. We graphically show this expansion for a typical, non-engineered PDC in Fig. \ref{fig:pdc_calculation_general}(a).

From Eqs. (\ref{eq:pdc_state}) and (\ref{eq:decomposition}), we obtain
\begin{equation}
    \ket{\psi}_\mathrm{PDC} =
    \sum_{k=0}^\infty\sqrt{\lambda_k}\ket{A_k,B_k}
    \label{eq:pdc_state_tm}
\end{equation}
where we used again the broadband mode operators from Eq. (\ref{eq:broadband_operator}). This expression shows that the PDC excites pairs of TM states $\ket{A_k}$ and $\ket{B_k}$ with a relative weight of $\sqrt{\lambda_k}$.

For the special case of a dispersion-engineered PDC that excites only a single pair of TMs (see, for instance \cite{URen:2005wb,Mosley:2008ir,Kuzucu:2008gv,Shi:2008tl,Eckstein:2011wl,Harder:2013hk}), the state from Eq. (\ref{eq:pdc_state_tm}) reduces to $\ket{\psi}_\mathrm{PDC}=\ket{A_0,B_0}$. This situation is shown in Fig. \ref{fig:pdc_calculation_general}(b). In this case, by detecting the photon created in one channel, one heralds the single-photon state in the other channel in a known, pure TM. We note, however, that this is not sufficient for generating resource states for QIS applications. On the one hand, the general PDC state has an inadequate structure, because the number of TMs in the state cannot be precisely controlled. On the other hand, the single-TM state does not constitute an entangled resource state, which is a necessary requirement for different QIS applications.

\subsection{Coherent manipulation of the TM structure of single-photon states}
A major requirement for realizing QIS with TMs is the coherent manipulation of a state in the TM basis. This can be achieved by deploying so-called quantum pulse gates (QPGs) \cite{EcksteinA:2011vg,Brecht:2011hz,Reddy:2014bt,Reddy:2015jb}. Note that although we restrict our discussions to three-wave mixing implementations of QPGs here, all results can be generalized to four-wave mixing. The underlying physical process of a QPG based on three-wave mixing is dispersion-engineered sum-frequency generation (SFG) inside a nonlinear optical waveguide, where one photon from an ultrafast pump pulse and a ``red'' quantum signal fuse into a ``green'' converted output photon. Here, red and green describe two well-separated frequency bands, for instance $1535\,$nm (red) and $557\,$nm (green), respectively \cite{Brecht:2014eg}. An adaption of this approach for use with continuous-variable quantum states has been proposed in \cite{Averchenko:2014dn}. In four-wave mixing implementations, two non-degenerate pump pulses are used, which facilitate smaller frequency shifts of single photons as compared to using three-wave mixing \cite{Mechin:2006du,McGuinness:2011fv,McGuinness:2010ja}. 

An ideal QPG that is mode matched to the TMs of the source as defined above acts on an arbitrary single-photon input state $\ket{\psi}_\mathrm{in}$ of the form Eq. \ref{eq:TM_qudit} according to
\begin{equation}
    \ket{\psi}_\mathrm{out} = \hat{Q}_i^{(\eta)}\ket{\psi}_\mathrm{in}
\end{equation}
with 
\begin{equation}
    \begin{split}
    \hat{Q}_i^{(\eta)} &= \mathbb{1} - \ket{A_i}\bra{A_i} - \ket{C}\bra{C}\\
     &+ \cos\theta_i\left(\ket{A_i}\bra{A_i} + \ket{C}\bra{C}\right)\\
     &+ \sin\theta_i\left(\ket{C}\bra{A_i} - \ket{A_i}\bra{C}\right).
     \label{eq:qpg_operator}
    \end{split}
\end{equation}

\begin{figure}
    \centering
    \includegraphics[width=\linewidth]{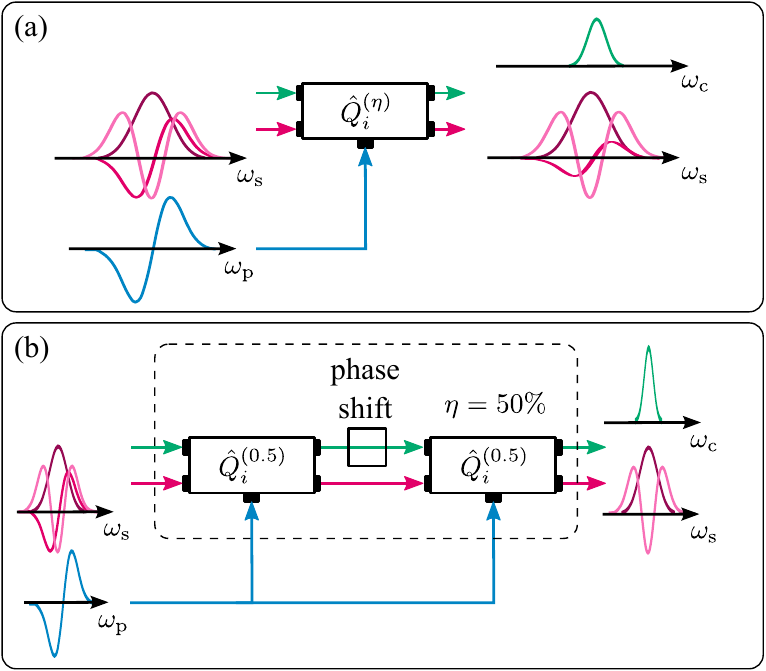}
    \caption{(a) Schematic of the QPG operation. The shape of the blue
      pump pulse selects one TM from the ``red'' input signal and
      converts it to the ``green'' output with an efficiency of
      $\eta$. All other signal TMs are completely transmitted. The
      index $i$ labels the addressed input TM. (b) A
      Mach-Zehnder/Ramsey like configuration of two successive QPGs
      with an efficiency of 50\% each overcomes the time-ordering
      limitations of a single QPG and facilitates the selection and
      conversion of a single TM with an efficiency of 100\%.}
    \label{fig:qpg_schematics}
\end{figure}

The cosine term preserves either of the two states of interest, while the sine term ``swaps'' them with efficiency $\sin^2(\theta_i)$. The first three terms enforce unitarity. This expression is a family of unitary transformations on the single-photon state space comprised of two non-overlapping subspaces (here, frequency bands): one spanned by the TM states $\ket{A_j}$, and a single TM state $\ket{C}$ occupying the other. It has an elegant interpretation: the QPG acts as a quantum mechanical beamsplitter, which operates on TMs instead of polarization or spatial modes. As detailed in \cite{Raymer:2010df,Brecht:2011hz}, the blue pump pulse spectrum $\alpha(\omega)$ defines the targeted ``red'' input TM state $\ket{A_i}$ that is selected and converted to the ``green'' output state $\ket{C}$ with an efficiency given by $\eta=\sin^2(\theta_i)$. Note that the QPG can also select superpositions of TM states, when the pump pulses are shaped accordingly. The parameter $\theta_i$ describes the strength of the QPG operation and can be tuned with the pump pulse energy, although the shapes of the ``red'' and ``green'' modes will change slightly for different values of $\theta_i$, due to time-ordering corrections \cite{Reddy:2013ip,Christ:2013fg,Quesada:2014gn} (i.e. the input and output TMs are not identical). For genuine QPG operation, $\theta_j=0$ for $j\neq i$; that is, all TMs that are not addressed are completely transmitted. This situation is sketched in Fig. \ref{fig:qpg_schematics}(a).

From Eq. (\ref{eq:qpg_operator}) we see two things. First, the QPG converts any targeted input state $\ket{A_i}$ into the same output state $\ket{C}$. This is important in light of large network architectures, because it facilitates interference between formerly orthogonal TM states after the QPG operation. Second, the QPG can also be operated ``backwards''. In this case, it accepts one single input state $\ket{C}$, which is coherently reshaped to an arbitrary output TM state $\ket{A_i}$. This allows the treatment of the $\ket{C}$ frequency band as a buffer, or ``processing'' state space, and allows one to perform arbitrary linear operations on TM qudits that reside in the $\{\ket{A_j}\}$-space using combinations of QPGs, as will be shown below.

A measure to quantify the operation fidelity of a QPG is the so-called
\textit{temporal mode-selectivity} \cite{Reddy:2014bt} 
\begin{equation}
    S = \frac{\sin^4(\theta_i)}{\sum_{j=0}^\infty\sin^2(\theta_j)}\leq1,
\end{equation}
which measures the ratio between the squared conversion efficiency of the selected mode and the conversion efficiencies of all modes. A mode selectivity of $1$ characterizes perfect single-TM operation, whereas a mode selectivity of $0$ signifies a total absence of modal selectivity. 

It has been shown that the single-stage QPG realization from Fig. \ref{fig:qpg_schematics}(a) cannot exceed a mode selectivity of $S=0.85$ due to the effects of time ordering, which lead to a temporal multimode behavior at conversion efficiencies exceeding 90\% \cite{Christ:2013fg,Reddy:2013ip}. This limitation can be overcome by utilizing a two-stage Mach-Zehnder/Ramsey like setup of two successive QPGs with an efficiency of 50\% each, which are driven by the same pump pulse shape \cite{Reddy:2014bt,Reddy:2015jb}. We sketch this in Fig. \ref{fig:qpg_schematics}(b).

In the two-stage QPG a single photon in the target TM will be converted into an equal superposition of a ``green'' and a ``red'' mode by the first stage, and will then be coherently fully frequency shifted or back-converted in the second stage depending on an externally applied phase shift to the device. The non-target TM components of the photon will not participate in the interferometric conversion process due to their vanishingly small per-stage conversion efficiencies, and will effectively transparently pass through the device. The need for phase coherence across the two stages can be met by deriving the two pump pulses from the same master pulse. In a specific configuration \cite{Reddy:2014bt,Reddy:2015jb}, this method also eliminates the temporal distortion in the shapes of the ``red'' and ``green'' modes due to time-ordering effects, which enables the cascading of QPGs without the need for inter-QPG compensatory TM reshaping. Note that the overall operation of the two QPGs is again collectively described by Eq. (\ref{eq:qpg_operator}) and that we use the simplified sketch from Fig. \ref{fig:qpg_schematics}(a) for reasons of convenience from here on. Various overall efficiency values can now be achieved by tuning the interferometric phase shift in-between the two stages (Fig. \ref{fig:qpg_schematics}(b)) instead of changing the pump power. 

In a recent experiment, the implementation of a single-stage QPG with an TM selectivity of $80\%$ at a conversion efficiency of $\eta=87\%$ when operated at the single-photon level has been demonstrated \cite{Brecht:2014eg}. 

Note that  alternative approaches to TM-selective SFG are studied in \cite{Huang:2013tu,Kowligy:2014ga,Donohue:2015cd}, which forego group-velocity matching. Although potentially simpler from an experimental point of view, these approaches cannot generally reach high mode selectivities as defined above \cite{Reddy:2013ip}.

\begin{figure*}
    \includegraphics[width=\linewidth]{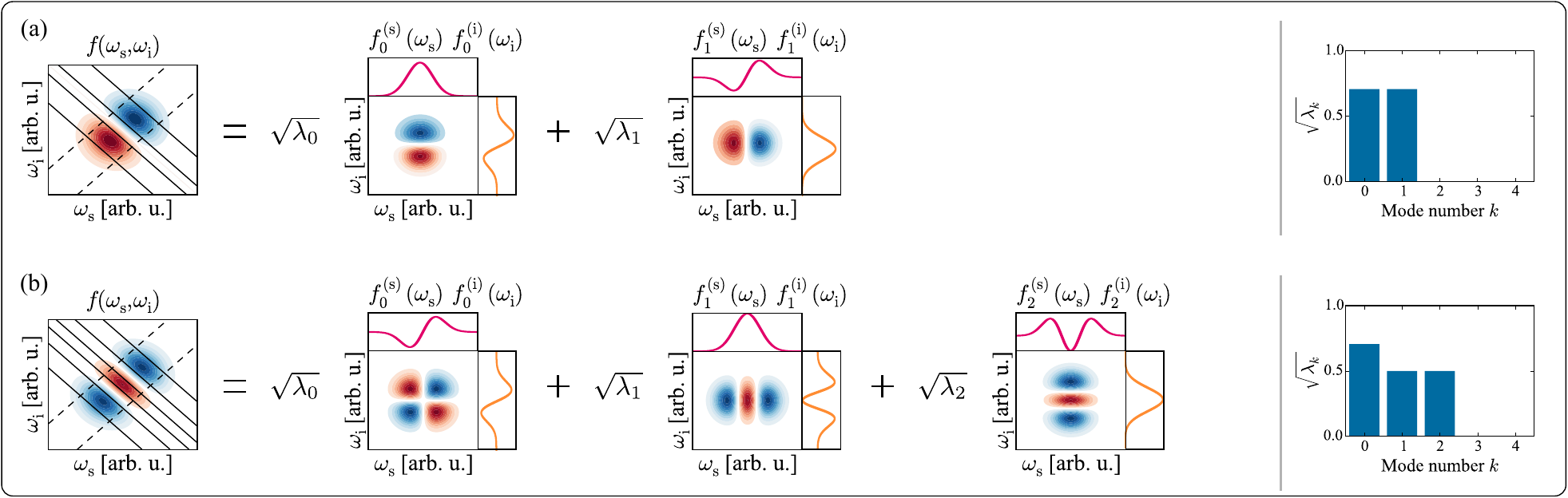}
    \caption{(a) When pumping a dispersion engineered PDC with a 
    $1^\mathrm{st}$ order Hermite-Gaussian pulse, the resulting JSA (left) has
    a negative part signified by the red color. Note that the pump
    envelope function is again denoted by solid black lines, whereas the
    phase-matching function is shown as dashed black lines. A decomposition of
    this JSA yields exactly two pairs of TMs (center) with similar expansion
    coefficients (right). Hence, the generated state is a TM Bell state. 
    (b) By further increasing the order of the Hermite-Gaussian pump, it is
    possible to successively add TM pairs to the generated state. This state
    features an extremely well-defined dimensionality, although the relative
    weights of the modes become unbalanced.}
    \label{fig:pdc_calculation_engineered}
\end{figure*}

\section{Completing the tool kit for a TM QIS framework}
In this section, we introduce the missing components, which enable our TM framework. In particular, these are the generation of TM states with an arbitrary, user-defined dimension and their verification using single-photon and photon-pair TM tomography. Thereafter, we show that ideal QPGs can be used to implement linear-optics single- and pair-photons quantum operations.

\subsection{TM engineering and TM Bell states}
Typical QIS applications require at least the faithful generation of Bell states. In the following, we demonstrate how this can be accomplished for TMs by combining in a very natural way a dispersion-engineered PDC with pulse-shaping techniques, which are well-established in the fields of ultrafast optics and coherent control (for a nice review see \cite{Monmayrant:2010gk}).

To this end, we consider shaped pump pulses with Hermite-Gaussian spectra given by
\begin{equation}
    \alpha(\os,\oi) = \frac{1}{\sqrt{n!\sqrt{\pi}2^n\sigma}}
    H_n\left(\frac{\Delta\omega}{\sigma}\right)
    \exp\left[-\frac{(\Delta\omega)^2}{2\sigma^2}\right].
\end{equation}
Here, $\Delta\omega=\omega_\mathrm{p}-\os-\oi$ is the frequency mismatch between the pump, signal and idler fields, $H_n(x)$ is a Hermite polynomial of order $n$ and $\sigma$ is the spectral $1/e$-width of the pump spectral intensity.

Fig. \ref{fig:pdc_calculation_engineered}(a) shows an engineered PDC that is driven by a $1^\mathrm{st}$ order Hermite-Gaussian pump pulse. The JSA decomposes into 
\begin{equation}
	f(\os,\oi) = \frac{1}{\sqrt{2}}\left(\fs_0(\os)\fid_0(\oi) +
	\fs_1(\os)\fid_1(\oi)\right).
\end{equation}
This result can be interpreted such that the PDC comprises exactly two pairs of TMs with equal excitation probability. Consequently, we write the generated photon-pair state as 
\begin{equation}
    \begin{split}
    \ket{\psi}_\mathrm{PDC}&\approx\frac{1}{\sqrt{2}}\left(
    \ket{A_0,B_0}+\ket{A_1,B_1}\right)=\\
    &\frac{1}{\sqrt{2}}\left(\ket{\gauss_\mathrm{s},\hermite_\mathrm{i}}+
    \ket{\hermite_\mathrm{s},\gauss_\mathrm{i}}\right),
    \end{split}
\end{equation}
where the graphical representation in the second line highlights the shapes of the individual signal and idler TMs. This state is a TM $\ket{\psi^+}$ Bell state, which is a fundamental resource for QIS applications. 

In Fig. \ref{fig:pdc_calculation_engineered}(b), we consider a $2^{\mathrm{nd}}$ order Hermite-Gaussian pump pulse. The decomposition of the resulting JSA shows that the generated state comprises exactly three TM pairs. Although the relative weights are not evenly distributed anymore, the dimensionality of the state is well-defined. Further increasing the order of the pump Hermite-Gaussian pulse successively adds additional TM pairs to the structure of the generated state. 

In this way it is possible to generate high-dimensional photonic states with an unprecedented degree of control. We emphasize again that all TMs ``live'' inside the same transverse spatial waveguide mode, which makes our approach exceptionally robust and guarantees experimental simplicity. 

\begin{figure}
    \centering
    \includegraphics[width=\linewidth]{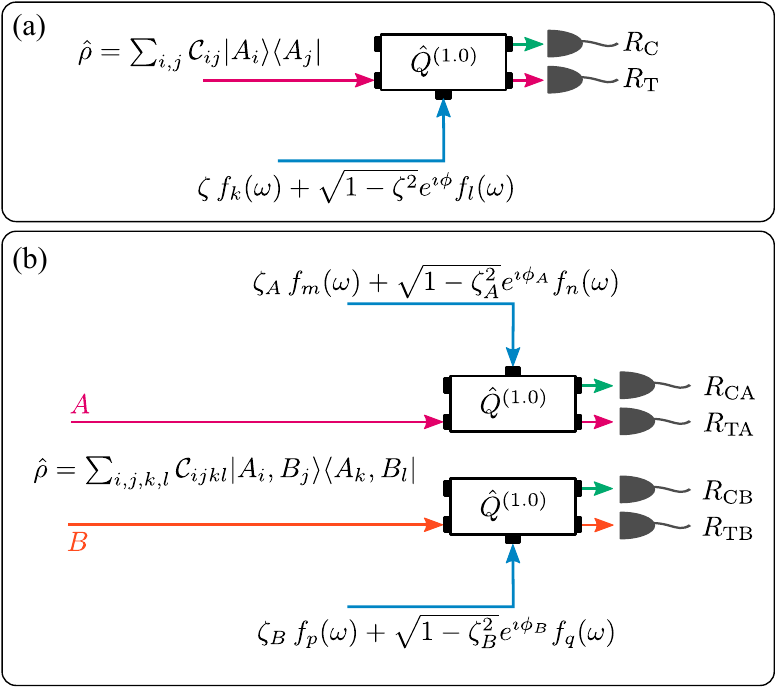}
    \caption{(a) TM state tomography of a single-photon state with density
    matrix $\hat{\rho}$. Both transmitted and converted output of the QPG are
    detected with single-photon detectors. (b) Generalized scheme for the
    TM tomography of a biphoton state. Photons ``1'' and ``2'' are sent to
    two different QPGs and the transmitted and converted outputs are detected
    with single photon detectors.}
    \label{fig:state_tomography}
\end{figure}

\subsection{Photon TM-state tomography}
With the ability to generate TM states with arbitrary dimension, the missing element to render a QIS framework based on TMs feasible is the verification of the state generation. To this end, we require TM state tomography, where the challenge is to retrieve the (complex-valued) entries of a quantum state's density matrix in a basis of TMs. This differs from polarization-state tomography because of the higher dimensionality of the TM-state space. For an arbitrary single-photon state, the density matrix is given by

\begin{equation}
	\hat{\rho}=\sum_{i,j}\mathcal{C}_{ij}\ket{A_i}\bra{A_j},
\end{equation}
with associated TMs $\{f_i(\omega)\}$. This state can be analyzed with a QPG, which selects a coherent superposition of TMs given by $\zeta\, f_k(\omega)+\sqrt{1-\zeta^2}e^{\imath\phi} f_l(\omega)$, where $\zeta\in[0,1]$, as shown in Fig. \ref{fig:state_tomography}(a). This function is defined by the shape of pump pulse the QPG is ``programmed'' with. Detecting both the converted output and the transmitted light with single photon detectors, we measure the average converted count rates $R_\mathrm{C}$ and $R_\mathrm{T}$ respectively, which are related to elements of the input density matrix by
\begin{equation}
	\begin{split}
	\frac{R_\mathrm{C}}{R_\mathrm{C}+R_\mathrm{T}} &=
	\zeta^2\mathcal{C}_{kk} + (1-\zeta^2)\mathcal{C}_{ll} \\
	& + 2\mathfrak{Re}\left[\zeta\sqrt{1-\zeta^2}e^{\imath\phi}
	\mathcal{C}_{lk}\right].
	\end{split}
\end{equation}

From this expression we see that for $\zeta=0$ and $\zeta=1$, we directly obtain $\mathcal{C}_{kk}$ and $\mathcal{C}_{ll}$, respectively. To retrieve the complex coefficient $\mathcal{C}_{lk}$, we set $\zeta=\frac{1}{\sqrt{2}}$ and evaluate the counts for $\phi=0$ and $\phi=\frac{\pi}{2}$. By extension, we also obtain $\mathcal{C}_{kl}$ and thus a complete subset of matrix coefficients of the density matrix $\hat{\rho}$. In this way, the complete density matrix or an experimentally feasible subset thereof can be sampled. It is important to note that any chosen portion of the density matrix can be ``directly'' measured in this way without reconstructing the entire state. This is true only for a QPG that can achieve unit selectivity, although without high selectivity, the elements can still be found up to an unknown normalization constant. This would necessitate measuring the entire matrix (or making small-magnitude assumptions about the unmeasured coefficients).

This procedure can be generalized to certain biphoton states as sketched in Fig. \ref{fig:state_tomography}(b). A general two-photon state in two different spatial modes (with photon labels $A$ and $B$) may be expressed in two sets of TM bases as
\begin{equation}
    \hat{\rho}=\sum_{i,j,k,l}\mathcal{C}_{ijkl}\ket{A_i,B_j}
    \bra{A_k,B_l}.
\end{equation}

The two photons are analyzed with two separate QPGs, which select TMs given by $\zeta_A\,f_m(\omega)+\sqrt{1-\zeta_A^2}e^{\imath\phi_A}f_n(\omega)$ and $\zeta_B\,f_p(\omega)+\sqrt{1-\zeta_B^2}e^{\imath\phi_B}f_q(\omega)$, respectively. Then we employ four single-photon detectors labeled CA, TA, CB, and TB, as shown in Fig. \ref{fig:state_tomography}(b). We can then measure coincidence rates between pairs of detectors (say between CA and CB, denoted by $R_{CA,CB}$, and so on). The following expression of such coincidence rates
\begin{equation}
    \frac{R_{CA,CB}}{R_{CA,CB}+R_{CA,TB}+R_{TA,CB}+R_{TA,TB}}
\end{equation}
can be expressed in terms of the biphoton density matrix elements thusly
\begin{widetext}
\begin{equation}
    \begin{split}
        \zeta_A^2 & \zeta_B^2\mathcal{C}_{mppm}
        +(1-\zeta_A^2)(1-\zeta_B^2)\mathcal{C}_{nqqn}
        +\zeta_A^2(1-\zeta_B^2)\mathcal{C}_{mqqm}
        +(1-\zeta_A^2)\zeta_B^2\mathcal{C}_{nppn}\\
        & 2\mathfrak{Re}\left[
        e^{\imath\phi_A}\zeta_A\sqrt{1-\zeta_A}\left(
            \zeta_B^2\mathcal{C}_{mppn}+(1-\zeta_B^2)\mathcal{C}_{mqqn}
            \right)
        +e^{\imath\phi_B}\zeta_B\sqrt{1-\zeta_B}\left(
            \zeta_A^2\mathcal{C}_{mpqm}+(1-\zeta_A^2)\mathcal{C}_{nqqn}
            \right)\right.\\
        & \left.+\zeta_A\zeta_B\sqrt{1-\zeta_A}\sqrt{1-\zeta_B}\left(
            e^{\imath(\phi_A+\phi_B)}\mathcal{C}_{mpqn}+
            e^{\imath(\phi_A-\phi_B)}\mathcal{C}_{mqpn}\right)
        \right]
    \end{split}
\end{equation}
\end{widetext}

Cycling through the parameter space $(\zeta_{1,2}, \phi_{1,2})\in \{(1,-), (0,-), (\frac{1}{\sqrt{2}}, 0), (\frac{1}{\sqrt{2}}, \frac{\pi}{2})\}$ as well as varying the indices $(m,n,o,p)$ will reveal any desired set of coefficients from the two-photon density matrix.

\section{QIS Applications}
In this section, we combine the different building blocks to detail several QIS applications, which can be realized in the TM framework and highlight its versatility. We first consider photon TM purification and TM reshaping, then move on to quantum communication scenarios and conclude with considerations on single-qubit gate operations and cluster-state generation. Note that we will discuss the technical challenges that have to be faced when implementing these applications in detail in the following section. 

\begin{figure}
    \centering
    \includegraphics[width=\linewidth]{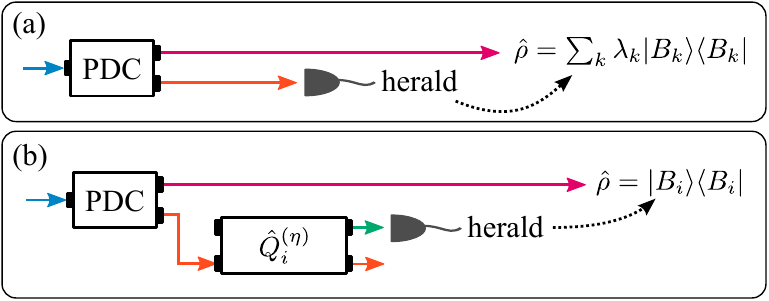}
    \caption{(a) Non mode-selective detection of one PDC photon generally
    projects its sibling into a mixed state. (b) Deploying a QPG to herald a
    single TM yields a pure heralded broadband photon at the cost of a lowered
    heralding rate.}
    \label{fig:detection}
\end{figure}

\subsection{Photon TM ``purification''}
Let us consider an application, which requires either a photon-pair at very specific wavelengths or a choice of nonlinear material, such that it is not possible to directly implement a dispersion-engineered PDC source which generates only a single pair of TMs, but instead a general PDC state as sketched in Fig. \ref{fig:pdc_calculation_general}(a). In this case, people typically resort to spectrally narrow intensity filtering to facilitate the heralding of approximately pure single photons, thus discarding the greater portion of the generated photon pairs \cite{Riedmatten:2003cl,Kaltenbaek:2006ch}. Our TM tool kit provides a more efficient and elegant solution to this problem, which additionally facilitates the heralding of genuinely pure broadband single photons from a correlated source such as shown in Fig. \ref{fig:pdc_calculation_general}(a).

We assume the general PDC state from Eq. (\ref{eq:pdc_state_tm}) and detect one of the photons, say photon $A$, with an unfiltered single-photon detector as sketched in Fig. \ref{fig:detection}(a). This heralds photon $B$ with a reduced density matrix that is given by
\begin{equation}
	\hat{\rho}_B=\sum_{k=0}^{\infty}\lambda_k\ket{B_k}\bra{B_k},
\end{equation}
which is generally a mixed state with purity $\mathcal{P}=\sum_k\lambda_k^2$.

On the other hand, we can send photon $A$ to a QPG, which acts as a complex spectral-amplitude shape ``filter'' that selects a single TM $\fs_i(\os)$ with efficiency $\eta$, and detect only the converted output. In this case, a successful detection heralds photon $B$, which is in a pure state with corresponding density matrix 
\begin{equation}
	\hat{\rho}_B=\ket{B_i}\bra{B_i},
\end{equation}
as sketched in Fig. \ref{fig:detection}(b) \cite{EcksteinA:2011vg}. Note that this ``purification'' comes at the cost of a lower heralding rate, which is reduced by the factor $\lambda_i$. Still, the advantage is that a photon in a desired TM can be created, rather than simply a spectrally filtered photon. 

As a side remark, although we restricted our analysis to photon-pair states, the TM framework can directly be applied to continuous variable states. In this context, a particularly important non-Gaussian operation is TM-selective photon subtraction from a multimode state, which is required for entanglement distillation \cite{Eisert:2002wb}. It is based on the same operation as the photon TM ``purification'', but uses a QPG that is intentionally operated at very low conversion efficiency \cite{Averchenko:2014dn}.

\subsection{Single photon TM reshaping}
\begin{figure}
    \centering
    \includegraphics[width=\linewidth]{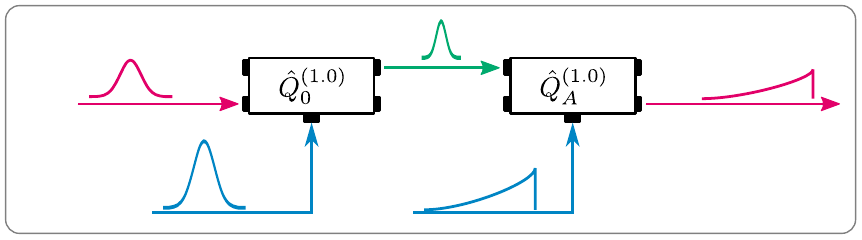}
    \caption{TM reshaping of a single photon. A QPG first converts the ``red''
	single photon to the ``green'' channel. A second QPG then reshapes the
	photon during back-conversion.}
    \label{fig:tm_reshaping}
\end{figure}
Large-scale networks require an efficient interfacing between distinct nodes. For different photon sources, this means that the photons have to be made indistinguishable. For coupling photons to solid-state systems, this means that the TM of the photons has to match the acceptance TM of the system. In both cases, a coherent TM reshaping of the photons is preferable to other filtering operations, since the latter introduce prohibitive losses. In Fig. \ref{fig:tm_reshaping}, we sketch a TM reshaper: A first QPG converts the ``red'' input photon -- which we implicitly assume to be pure and thus TM single-mode -- to the ``green'' channel; A second QPG is then used to back-convert the photon to the ``red'' channel. However, here we match the shape of the bright pump pulse to the required TM and by this reshape the photon. Note that the reshaped mode does not have to be a mode from the original photon TM basis, which is indicated by the label $A$ (as opposed to a numeric label) of the QPG operation in the figure. The complete reshaping operation can then be written as
\begin{equation}
	\begin{split}
		\ket{\psi}_A &= \hat{Q}_A^{(1.0)}\hat{Q}_0^{(1.0)}\ket{A_0}\\
		 &= \hat{Q}_A^{(1.0)}\ket{C} = -\ket{A_A},
	\end{split}
\end{equation}
where we assumed the original photon to be in the TM state $\ket{A_0}$ and the overall phase of the output state can be neglected. The operators $\hat{Q}_i^{(1.0)}$ are the QPG operators from Eq. (\ref{eq:qpg_operator}). In principle, arbitrary reshaping is possible in this way. Note that a reshaping of the ``green'' TM can be realized by tailoring the phasematching function of the QPG \cite{Branczyk:2011uw,BenDixon:2013tk,Dosseva:2014wr}. In this way, an adapted interface between photons at telecommunication wavelengths and specific quantum memories can be realized with a single QPG.

\begin{figure}
    \centering
    \includegraphics[width=\linewidth]{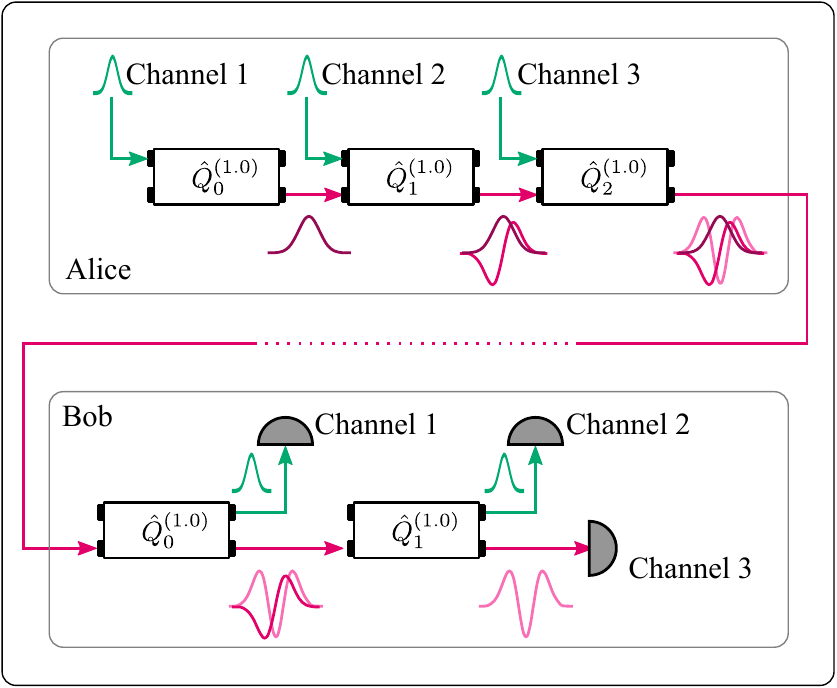}
    \caption{In a TM multiplexing scenario, Alice uses orthogonal TMs as
      independent channels, which are sent to Bob in one single
      physical fiber. He de-multiplexes the channels with QPGs and
      reads out the information. The QPGs are being employed as
        TM multiplexers (Alice) and demultiplexers (Bob) on a
        single-mode optical channel.}
    \label{fig:quantum_communication}
\end{figure}

\subsection{Quantum communication}
Another important aspect of QIS is quantum communication (QC), where quantum information is transmitted between distant parties, by convention called Alice and Bob. To this end, information has to be encoded at Alice's location and decoded and read out at Bob's location. Deploying the aforementioned devices and methods, a QC system based on TMs can be readily set up.

Here, we discuss two approaches to realizing this. The first approach utilizes different TMs as different communication channels and thus relies on TM multiplexing. Note that in this approach, information is not encoded in the TMs but in another degree of freedom, for instance the polarization. The second approach directly encodes the information in arbitrary superpositions of single-photon TMs, and thereby can implement genuine high-dimensional QC.

The use of TMs for channel multiplexing would be distinguished from conventional time- or frequency-based optical multiplexing, which use either separated short pulses or narrow spectral windows to define different information channels. Such schemes have recently been proposed in the general context of QIS as well \cite{Nunn:2013kf, Humphreys:2014ce}. However, they are not based on genuinely field-orthogonal modes, which translates to a lower ``packing density'' of signal channels in time-frequency space to ensure approximate orthogonality. A fundamental advantage of our TM approach is that it is intrinsically based on genuinely field-orthogonal wave-packet modes, which provide in-principle zero cross talk between mode channels, while densely packing these modes in time-frequency phase space. 

\begin{figure*}
    \centering
    \includegraphics[width=\linewidth]{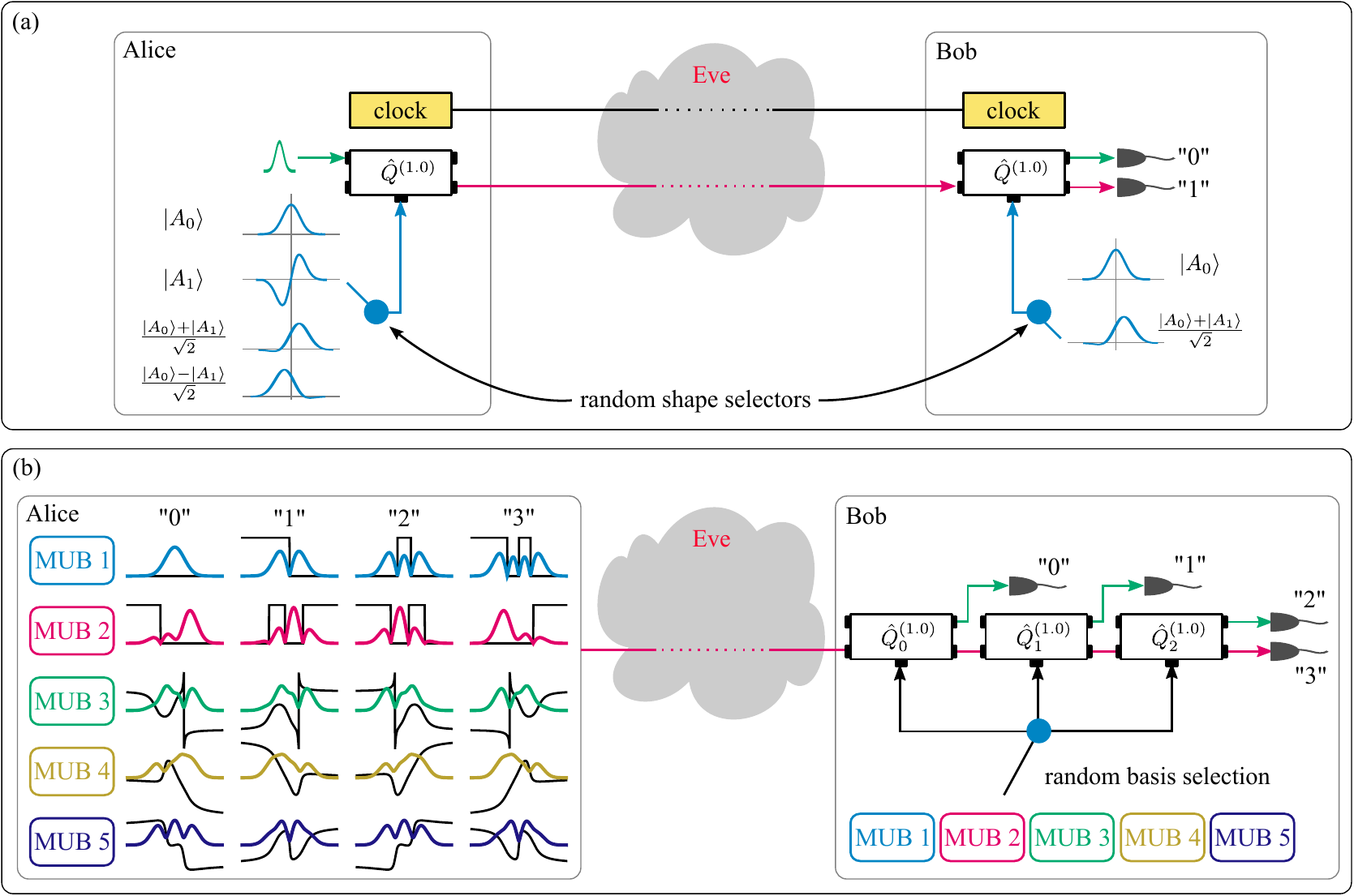}
    \caption{(a) Implementation of the BB84
      QKD protocol with TMs. Alice randomly prepares one of four
      possible basis states and sends it to Bob, who randomly measures
      in one of two MUBs. The two outputs of Bobs QPG correspond to
      ``0'' and ``1''. (b) Generalized BB84 in a four-dimensional encoding
	  scheme. Alice randomly prepares one of the 20 possible basis states. Bob
	  chooses randomly one of the five MUBs to measure. Note that in this case
	  he requires three QPGs to fully resolve the four possible basis states
	  of each MUB.}
    \label{fig:quantum_communication_2}
\end{figure*}

In QC, for a TM multiplexing a scheme to work, add/drop functionality is essential. Using the QPG, both operations can be implemented as sketched in Fig. \ref{fig:quantum_communication}(a). On Alice's side, a succession of QPGs adds different channels to the communication line. This is possible due to the TM-selective operation of the QPG, which reshapes the ``green'' input $\fc(\omega)$ into the desired ``red'' TM $\fs_i(\omega)$. At the same time, the existing ``red'' TMs $\fs_j(\omega)$ with $j\neq i$ are not affected. Note that this operation mode of the QPG has been referred to as quantum pulse shaper earlier \cite{Brecht:2011hz}. After transmission, Bob deploys a cascade of QPGs to de-multiplex the different channels into separate ports, from which the information is read out \cite{Huang:2013tu}.

The second approach, high-dimensional QC, is appealing in light of quantum key distribution (QKD) applications, where the goal is to establish a secure encryption key between Alice and Bob. Deploying TMs, the implementation of a generalized BB84 protocol \cite{Bennett:1984wv} becomes possible. To clarify this procedure, we first sketch the realization of the original BB84 protocol using two TM MUBs instead of polarization in Fig. \ref{fig:quantum_communication_2}(a). Alice randomly prepares one of the four possible basis states with a QPG and sends it to Bob. Bob in turn randomly chooses the measurement basis of his QPG and directly detects both output ports, which then correspond to `$0$' and `$1$'. Thereafter, Alice and Bob publicly announce their preparation and measurement bases and keep only those events when both coincide. Sacrificing a part of the so retained key, Alice and Bob can uncover an eavesdropper by the $25\%$ error he or she inevitably introduces.

This scheme is readily extended to $d$ dimensions. We illustrate this for the case of $d=4$, which is depicted in Fig. \ref{fig:quantum_communication_2}(b). In this case, five MUBs and thus a total of 20 possible basis states exist, from which Alice randomly chooses one. The four basis states of each MUB now encode logical ``$0$'' to ``$3$''. In the figure, we used the first four Hermite-Gaussian pulses as the ``mother'' basis from which ``daughter'' MUBs are created. Again, Alice transmits the chosen state to Bob who performs the readout in a randomly chosen basis. Note however that Bob now requires three QPGs to completely separate the four basis states of the MUBs. More generally, Bob requires $d-1$ QPGs for a basis of size $d$. There are two major advantages to high-dimensional encoding schemes in QC. On the one hand, high-dimensional encoding facilitates a higher information capacity per photon, and thus leads to a reduction in the overall number of required photons. On the other hand, it has been shown that high-dimensional encoding can increase the security of quantum key distribution, due to a larger error that is introduced by a potential eavesdropper when intercepting the transmission \cite{BechmannPasquinucci:2000ug,Cerf:2002fp}.

\subsection{Quantum computation}
In this section, we discuss two routes towards quantum computation enabled by the completion of the TM tool kit. First, we consider linear optical quantum computation (LOQC), where TM qubits propagate through a linear optical network and are subject to single- and two-qubit operations, which define the computation algorithm. Then, we investigate cluster-state quantum computation, where multiple TM qubits are fused in a specific way to create a graph state with a tailored entanglement structure. Then, measurements of the nodes (photons) of the cluster state implement the computation algorithm, the result of which can be read out from the remaining nodes. Although universal photonic quantum computation is beyond today's technological capabilities \cite{Li:2015ue}, the required operational building blocks can be realized with TMs.

Since in this paper we focus on three-wave mixing implementations of QPGs, we are effectively restricted to one single ``green'' output TM state $\ket{C}$, though we allow for a complete set of ``red'' input TM states $\ket{A_i}$. Consequently, the input states are treated as the primary qudit information ``register'' space, and the output channel will play the role of a ``processing'' space. Note that this behavior gives rise to the question whether QPGs are sufficient to realize all of the necessary operations for quantum computation. We will show in the following that they are.

\subsubsection{LOQC}
In LOQC, deterministic two-qubit operations are provably impossible. However, arbitrary single qubit operations can be implemented with a combination of QPGs. For this, we require two special cases of the QPG operation from Eq. (\ref{eq:qpg_operator}). First, a QPG with a conversion efficiency of $100\%$, and second, a QPG with a conversion efficiency of $50\%$. They are represented by operators
\begin{equation}
    \begin{split}
    \hat{Q}_i^{(1.0)} &= \mathbb{1} - \ket{A_i}\bra{A_i} - \ket{C}\bra{C}\\
          &+ \ket{C}\bra{A_i} - \ket{A_i}\bra{C},
     \label{eq:qpg_operator_100}
    \end{split}
\end{equation}
and
\begin{equation}
    \begin{split}
    \hat{Q}_i^{(0.5)} &= \mathbb{1} - \ket{A_i}\bra{A_i} - \ket{C}\bra{C}\\
     &+ \frac{1}{\sqrt{2}}\left(\ket{A_i}\bra{A_i} + \ket{C}\bra{C}\right)\\
     &+ \frac{1}{\sqrt{2}}\left(\ket{C}\bra{A_i} - \ket{A_i}\bra{C}\right).
     \label{eq:qpg_operator_50}
    \end{split}
\end{equation}

\begin{figure}
    \centering
    \includegraphics[width=\linewidth]{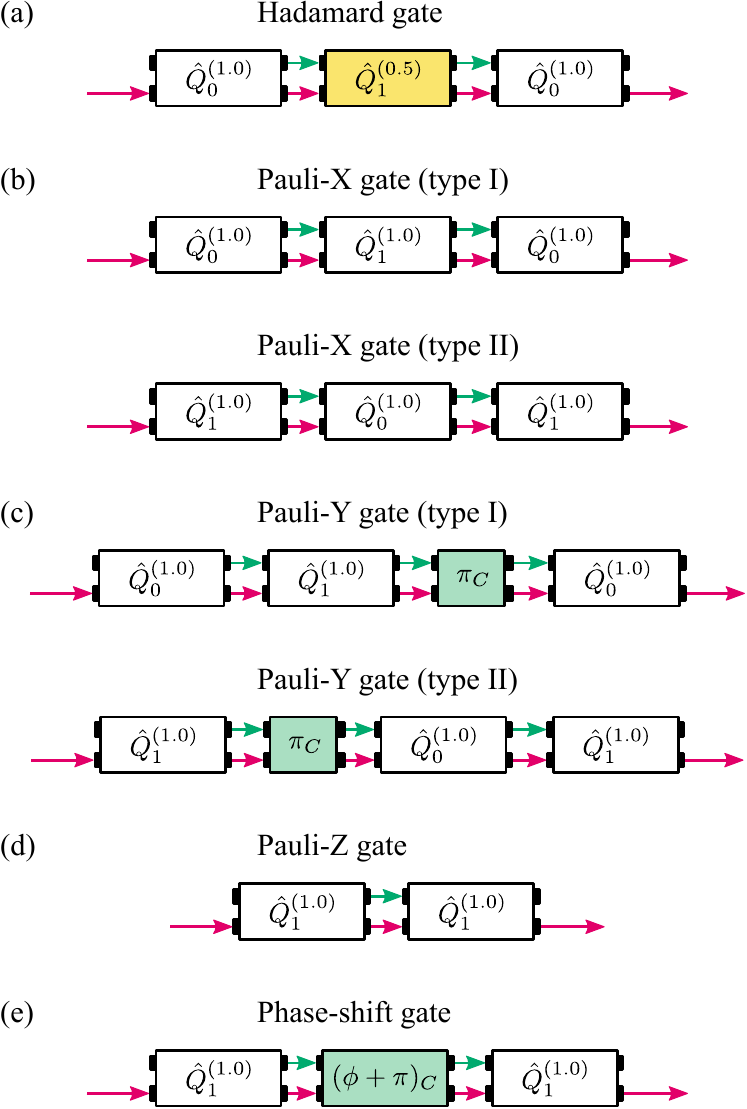}
    \caption{Implementation of single-qubit gates for LOQC using QPGs with
    $100\%$ conversion efficiency (white boxes), QPGs with $50\%$ conversion
    efficiency (yellow boxes) and phase shifts of the green $\ket{C}$ TM (green
    boxes). Note that both the Pauli-X gate and the Pauli-Y
    gate have two possible experimental implementations, which differ in the
    order in which the red TMs $\fs_0(\omega)$ and $\fs_1(\omega)$ are
    addressed.}
    \label{fig:loqc_operators}
\end{figure}

In Fig. \ref{fig:loqc_operators}, we show how these operations driven by the proper pump shapes can be sequentially combined with channel-dependent phase shifts, which shift the phase only in the ``green'' processing space and are shown as green boxes, to implement the following single-qubit operations (up to an overall phase) on the $\{\ket{A_0}, \ket{A_1}\}$ space:
\begin{itemize}
    \item[(a)] Hadamard gate:
    \begin{equation}
        \hat{H}=\frac{\ket{A_0}+\ket{A_1}}{\sqrt{2}}\bra{A_0}+
        \frac{\ket{A_0}-\ket{A_1}}{\sqrt{2}}\bra{A_1}
    \end{equation}
    \item[(b)] Pauli-X gate (type I, II):
    \begin{equation}
        \hat{X}=\ket{A_1}\bra{A_0}+\ket{A_0}\bra{A_1}
    \end{equation}
    \item[(c)] Pauli-Y gate (type I, II):
    \begin{equation}
        \hat{Y}=-\imath\ket{A_1}\bra{A_0}+\imath\ket{A_0}\bra{A_1}
    \end{equation}
    \item[(d)] Pauli-Z gate:
    \begin{equation}
        \hat{Z}=\ket{A_0}\bra{A_0}-\ket{A_1}\bra{A_1}
    \end{equation}
    \item[(e)] Phase-shift gate:
    \begin{equation}
        \hat{\phi}=\ket{A_0}\bra{A_0}+e^{\imath\phi}\ket{A_1}\bra{A_1}
    \end{equation}
\end{itemize}

These realizations rely on only two different pump shapes, corresponding to the ``red'' TMs $\fs_0(\omega)$ and $\fs_1(\omega)$, which encode the logical ``0'' and ``1''. The phase-shift gate can be simplified, if the phase $(\phi+\pi)$ is imprinted onto one of the two pump pulses. Then, the channel dependent phase shift can be omitted.

Note that the ``green'' channel is used only internally, whereas the input and output channels are the ``red'' TMs. This greatly reduces the challenge of maintaining phase relations between different frequency bands. It also eliminates the phase-coherence requirement for pump pulses across different red-channel-to-red-channel single-qubit gates, only requiring it for pump pulses internal to any given single-qubit gate. Additionally, the sequential steps can in principle be fabricated in monolithic devices, which promises a compact and robust implementation with building blocks that are well-suited to be used in integrated networks.

We also emphasize that, in a manner similar to \cite{Reck:1994dz}, any single \textit{qudit} operation can be realized with a concatenation of the single qubit operations outlined in this section. Then, the pump shapes have to be chosen such that the single qubit gates operate on every two-dimensional subspace of the qudit space successively.

\subsubsection{Cluster state quantum computation}
Finally, we consider the generation of discrete variable cluster states based on TMs. To efficiently grow cluster states from a supply of resource Bell pairs, we require several operations. Assuming that we already have a stock of linear cluster states which we want to merge into two-dimensional cluster states, we need local Hadamard transformations and projective measurements \cite{Gilbert:2006ba}. We have already shown how these can be implemented with TMs. More important is the ability to generate linear cluster states from Bell pairs. In order to do so, we have to rely on qubit fusion. A general method which facilitates this for polarization qubits has been introduced by Browne and Rudolph \cite{Browne:2005dd}, where it was referred to as \textit{Type-I fusion}. Here, we adapt this scheme to operate on TM qubits as defined in Eq. (\ref{eq:tm_qubit}).

\begin{figure}
\centering
\includegraphics{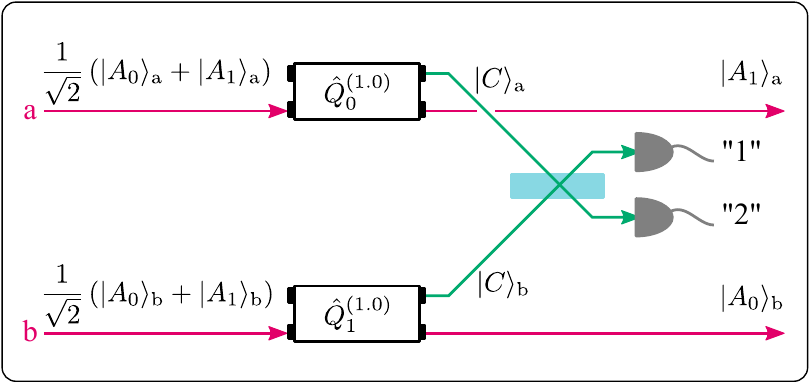}
\caption{Two TM qubits in spatial beams a
  and b can be fused with two QPGs, which select different ``red'' TM
  components from the qubits and selectively frequency-convert
  them. Then, the ``green'' outputs of the QPGs are interfered at a
  50/50 beamsplitter (blue rectangle) and detected with detectors
  ``1'' and ``2''. For more information, see the text.}
\label{fig:qubit_fusion}
\end{figure}

Two qubits in spatial beams a and b are sent to two QPGs as sketched in Fig. \ref{fig:qubit_fusion}. The QPGs implement the operation $\hat{Q}^{(1.0)}_0$ on qubit a and $\hat{Q}^{(1.0)}_1$ on qubit b, respectively. This means, that the ``red'' TMs $\fs_{0,\mathrm{a}}(\omega)$ and $\fs_{1,\mathrm{b}}(\omega)$ are converted to the green TMs $\fc(\omega)_\mathrm{a,b}$. The two ``green'' channels are interfered on a balanced beamsplitter behind the QPGs to erase any distinguishing information and the beamsplitter output ports are detected by detectors ``1'' and ``2''. The successful detection of a single ``green'' photon heralds the successful qubit fusion operation, which can be written in terms of Kraus operators
\begin{equation}
    \hat{O}_\mathrm{1,2} = \frac{1}{\sqrt{2}}\Big(
    \ket{A_0}_\mathrm{b}
    \bra{A_0}_\mathrm{a}\bra{A_0}_\mathrm{b} \mp
    \ket{A_1}_\mathrm{b}\bra{A_1}_\mathrm{a}\bra{A_1}_\mathrm{b}\Big),
\end{equation}
where the sign depends on whether detector ``1'' or ``2'' fires. The
state after a successful fusion is given by
\begin{equation}
    \ket{\psi}_\mathrm{fused} = \frac{1}{\sqrt{2}}\left(\ket{A_0}_\mathrm{b}
    \mp\ket{A_1}_\mathrm{a}\right),
\end{equation}
which, as expected, denotes again a qubit state. Note that the two parts of the fused qubit can be deterministically combined into a single spatial mode with the add/drop functionality of the QPG discussed in the context of quantum communication.

\section{Challenges}
In this section, we detail the challenges one faces when implementing QIS applications based on TMs.  While photonic quantum-information systems are ideal for serving as intermediary between memory, interaction, and detection resources, they come with known challenges. Most notably, the absence of any direct photon-photon interaction limits all-optical quantum information processing to nondeterministic logic gates \cite{Knill:2001vi}  or cluster-state measurement schemes \cite{Raussendorf:2001js}. When compared with optical-polarization or beam-path encoding of quantum information, the proposed TM encoding brings additional challenges, which need to be overcome in order to take advantage of the large in-principle benefits of using TMs for QIS: their relative immunity from channel dispersion and their compatibility with quantum memories in hybrid QIS systems, where efficient coupling into and out of disparate devices is highly dependent on temporal-mode matching which can, in principle, be achieved with TM reshaping. %

For this, the limiting factor is the bandwidth $\Delta\nu_\mathrm{PM}$ of the phasematching function of the QPG, which determines the minimal bandwidth of the reshaped TM. For the QPG presented in \cite{Brecht:2014eg} the spectrum of the ``green'' TM had a FWHM of $\Delta\lambda_\mathrm{g}=0.14\,$nm, corresponding to a bandwidth of $\Delta\nu_\mathrm{g}\approx135\,$GHz, which equals $\Delta\nu_\mathrm{PM}$ \cite{Brecht:2011hz}. We can calculate the narrowest possible phasematching bandwidth of a QPG based on a lithium niobate waveguide with uniform periodic poling. The maximum waveguide length is limited by the size of commercially lithium niobate crystals to around $L_\mathrm{max}\approx90\,$mm. Using this number, the resulting phasematching bandwidth is calculated to be $\Delta\nu_\mathrm{PM}\approx9.7\,$GHz, which is close to the maximum bandwidth of state-of-the-art quantum memories based on Raman interaction in warm Cs vapor of $9.2\,$GHz \cite{Reim:2011gr}. In addition, recent results on manipulating the phasematching function by manipulating the periodic poling pattern of waveguides \cite{Branczyk:2011uw,BenDixon:2013tk,Dosseva:2014wr} hold the promise for a future decrease of the effective phasematching bandwidth. Thus, deploying QPGs as interfaces between flying and stationary qubits is a realistic vision. %

An additional complication when interfacing flying and stationary qudits is the required multimode capability of the quantum memory. The Raman memory mentioned above, for instance, can store only a single TM \cite{Nunn:2007fi}. However, it has recently been shown that a concatenation of several Raman-type memories can overcome this limitation and store high-dimensional TM states \cite{Zheng:2015gm}. This result is a promising step towards the realization of high-dimensional hybrid quantum networks and facilitates the seamless integration of quantum memories into the TM framework. %

A further challenge is the achievable loss budget for a QIS application based on TMs. In this context, we highlight again that all TMs live inside the same spatial mode and thus feature low-loss transmission through standard single mode fibers. In addition, waveguide to fiber coupling with efficiencies exceeding 92\% has been demonstrated \cite{Alferness:1982vc}. Finally, waveguide propagation losses as low as $0.016\,$dB/cm in titanium-indiffused lithium niobate waveguides can be realized with state-of-the-art technology \cite{Luo:2015vz}. In total, we find a total insertion loss of roughly $1.0\,$dB for coupling from a fiber to a $90\,$mm long QPG and back to a fiber. In this case, the main losses arise from the fiber couplings. We note, that this challenge is not singular to the TM framework, but rather a challenge that the whole field of integrated quantum photonics has to face. Although current loss numbers are still prohibitively high, a significant increase in waveguide to fiber coupling efficiency can be expected in the coming years, which will alleviate this situation. %

Let us now focus on the realistically achievable number of TMs and thus the dimensionality of the accessible Hilbert space. With increasing mode order, the complexity as well as the spectral extent of TMs increases. Hence, the number of modes will be bounded, on the one hand, by the resolution of the deployed pulse shapers for pump pulses and, on the other hand, by the maximum spectral bandwidth of single-TM operation of the QPGs. For the demonstrator from \cite{Brecht:2014eg}, the maximum spectral bandwidth can be calculated to be around $25\,$nm for an input signal centered around $1550\,$nm. For larger bandwidths, the group-velocity dispersion inside the waveguide becomes non-negligible and the process is not TM single-mode anymore. Let us then assume that the TM states are generated with a PDC in a KTP waveguide as presented in \cite{Eckstein:2011wl,Harder:2013hk}. Then, the FWHM of the fundamental TM is around $5.0\,$nm. In this case, 10 TMs can be addressed with a selectivity in excess of $95\%$. A simple optimization of the PDC bandwidth and the length of the QPG waveguide increases this number to 20 TMs. Note that this is the limit only of the particular realization of a QPG based on lithium niobate waveguides. Investigating other materials with a more favorable group-velocity dispersion behavior can yield an even higher mode number. %

Concerning the resolution of state-of-the-art pulse shapers, we note that spatial light modulators with up to 4096 pixels are commercially available. Paired with proper imaging optics, these devices are capable of shaping TMs of order 100 with a fidelity of more than $99.9\%$. With respect to spatial light modulators, we also note their current limited switching speeds, which are typically in the order of few tens of kHz. These impose an upper limit on the switching speed of QIS applications. Again, this challenge does not only affect the TM framework, but also QIS based on transverse spatial modes, which also relies on SLMs as key elements. %

Now we consider the fidelity of the LOQG gate operations. In \cite{Reddy:2015jb}, Reddy et al. investigate the mode-selectivity of two-stage and multi-stage approaches to realizing QPGs. They found that in a two-stage QPG, the maximum selectivity is $S\approx98.46\%$, which translates to a maximum gate fidelity of around $95.4\%$, since every LOQC gate consists of three QPGs. This value cannot compete with requirements on fault-tolerant LOQC, but may facilitate small co-processing operations with only few gates. In addition, by increasing the number of stages in the QPG, the selectivity asymptotically approaches one. Thus there is a trade-off between the TM-selectivity and the total internal losses of a gate operation, which has to be evaluated in light of specific applications' requirements. %

Finally, let us turn our attention to the synchronization of the time-dependent, active components driven by shaped laser pulses in a TM framework. The timing requirements may be more severe when using TMs instead of other encoding bases, because the TM scheme relies essentially on temporal orthogonality, which is degraded under time jitter. To overcome this timing challenge over long-distance transmissions, we envision the use of weak coherent `pilot' pulses, which when amplified at the receiver can serve as a timing reference, a pump pulse, and a transmission-medium induced linear-dispersion compensator, all in one. %

In general, we find that, as with all burgeoning frameworks for optical QIS, the use of TMs will require significant investments in integrated device fabrication technology and timing electronics. TMs also share with other frameworks the need for efficient single-photon detection and lossless programmable optical routing. Ultimately, TM-based schemes might have to rely on performance gains from single-mode networkability and higher dimensionality, supplemented by their accommodation of broadband quantum memories, to outperform other optical QIS frameworks. %

\section{Conclusion}

We have shown that TMs of single-photon states form an appealing framework for QIS. Formally, they are comparable with transverse spatial field modes, but have distinct advantages over spatial modes: they are naturally compatible with waveguide technology, making them ideal candidates for integration into existing communication networks, and they are not affected by typical medium distortions such as linear dispersion, which renders them robust basis states for real-world applications. Still, TMs are of yet an underused resource for QIS. %

In this paper, we demonstrated that QIS based on TMs is feasible with current technology. We introduced a novel method for the generation of photon pair states comprising a user-defined number of TMs, which facilitates in particular the generation of TM Bell states. This method relies on the combination of dispersion-engineered PDC with classical pulse shaping for the pump pulses of the process. We then proposed TM tomography of single photon and photon-pair states as building blocks for a QIS framework based on TMs. %

Having established the necessary basis, we moved on to the implementation of QIS applications. With small photonic co-processing units embedded into large-scale hybrid quantum networks in mind, we first focussed on TM ``purification'' and TM reshaping. Thereafter, we discussed quantum communication based on TMs, where we presented two approaches: a TM multiplexing approach, where different TMs represented independent channels, and a high-dimensional TM QKD scenario, where the information was encoded into the order of the TMs. Finally, we demonstrated that any single qudit operation can be implemented with a succession of properly adjusted QPGs. We concluded the applications section with a scheme for TM cluster state generation which highlights the versatility of the TM framework. %

Finally, we discussed in detail technical challenges that must be faced when implementing QIS based on TMs. We expect that the introduction of this new framework will open novel research avenues in both fundamental and applied QIS. %

DVR and MGR were supported by the National Science Foundation through ENG-EPMD and PHYS-QIS. BB and CS acknowledge financial support by the Deutsche Forschungsgemeinschaft (DFG) via Sonderforschungsbereich TRR 142. %
\bibliography{/Users/bbrecht/work/tex_stuff/bibliography_abbr.bib}

\begin{thebibliography}{67}%
\makeatletter
\providecommand \@ifxundefined [1]{%
 \@ifx{#1\undefined}
}%
\providecommand \@ifnum [1]{%
 \ifnum #1\expandafter \@firstoftwo
 \else \expandafter \@secondoftwo
 \fi
}%
\providecommand \@ifx [1]{%
 \ifx #1\expandafter \@firstoftwo
 \else \expandafter \@secondoftwo
 \fi
}%
\providecommand \natexlab [1]{#1}%
\providecommand \enquote  [1]{``#1''}%
\providecommand \bibnamefont  [1]{#1}%
\providecommand \bibfnamefont [1]{#1}%
\providecommand \citenamefont [1]{#1}%
\providecommand \href@noop [0]{\@secondoftwo}%
\providecommand \href [0]{\begingroup \@sanitize@url \@href}%
\providecommand \@href[1]{\@@startlink{#1}\@@href}%
\providecommand \@@href[1]{\endgroup#1\@@endlink}%
\providecommand \@sanitize@url [0]{\catcode `\\12\catcode `\$12\catcode
  `\&12\catcode `\#12\catcode `\^12\catcode `\_12\catcode `\%12\relax}%
\providecommand \@@startlink[1]{}%
\providecommand \@@endlink[0]{}%
\providecommand \url  [0]{\begingroup\@sanitize@url \@url }%
\providecommand \@url [1]{\endgroup\@href {#1}{\urlprefix }}%
\providecommand \urlprefix  [0]{URL }%
\providecommand \Eprint [0]{\href }%
\providecommand \doibase [0]{http://dx.doi.org/}%
\providecommand \selectlanguage [0]{\@gobble}%
\providecommand \bibinfo  [0]{\@secondoftwo}%
\providecommand \bibfield  [0]{\@secondoftwo}%
\providecommand \translation [1]{[#1]}%
\providecommand \BibitemOpen [0]{}%
\providecommand \bibitemStop [0]{}%
\providecommand \bibitemNoStop [0]{.\EOS\space}%
\providecommand \EOS [0]{\spacefactor3000\relax}%
\providecommand \BibitemShut  [1]{\csname bibitem#1\endcsname}%
\let\auto@bib@innerbib\@empty
\bibitem [{\citenamefont {Kimble}(2008)}]{Kimble:2008if}%
  \BibitemOpen
  \bibfield  {author} {\bibinfo {author} {\bibfnamefont {H~J}\ \bibnamefont
  {Kimble}},\ }\bibfield  {title} {\enquote {\bibinfo {title} {{The quantum
  internet}},}\ }\href@noop {} {\bibfield  {journal} {\bibinfo  {journal}
  {Nature}\ }\textbf {\bibinfo {volume} {453}},\ \bibinfo {pages} {1023--1030}
  (\bibinfo {year} {2008})}\BibitemShut {NoStop}%
\bibitem [{\citenamefont {Aaronson}\ and\ \citenamefont
  {Arkhipov}(2011)}]{Aaronson:2011ho}%
  \BibitemOpen
  \bibfield  {author} {\bibinfo {author} {\bibfnamefont {Scott}\ \bibnamefont
  {Aaronson}}\ and\ \bibinfo {author} {\bibfnamefont {Alex}\ \bibnamefont
  {Arkhipov}},\ }\bibfield  {title} {\enquote {\bibinfo {title} {{The
  computational complexity of linear optics}},}\ }in\ \href@noop {} {\emph
  {\bibinfo {booktitle} {STOC '11}}}\ (\bibinfo  {publisher} {ACM},\ \bibinfo
  {address} {New York, New York, USA},\ \bibinfo {year} {2011})\ pp.\ \bibinfo
  {pages} {333--342}\BibitemShut {NoStop}%
\bibitem [{\citenamefont {Spring}\ \emph {et~al.}(2013)\citenamefont {Spring},
  \citenamefont {Metcalf}, \citenamefont {Humphreys}, \citenamefont
  {Kolthammer}, \citenamefont {Jin}, \citenamefont {Barbieri}, \citenamefont
  {Datta}, \citenamefont {Thomas-Peter}, \citenamefont {Langford},
  \citenamefont {Kundys}, \citenamefont {Gates}, \citenamefont {Smith},
  \citenamefont {Smith},\ and\ \citenamefont {Walmsley}}]{Spring:2013do}%
  \BibitemOpen
  \bibfield  {author} {\bibinfo {author} {\bibfnamefont {J~B}\ \bibnamefont
  {Spring}}, \bibinfo {author} {\bibfnamefont {B~J}\ \bibnamefont {Metcalf}},
  \bibinfo {author} {\bibfnamefont {P~C}\ \bibnamefont {Humphreys}}, \bibinfo
  {author} {\bibfnamefont {W~S}\ \bibnamefont {Kolthammer}}, \bibinfo {author}
  {\bibfnamefont {X.-M.}\ \bibnamefont {Jin}}, \bibinfo {author} {\bibfnamefont
  {M}~\bibnamefont {Barbieri}}, \bibinfo {author} {\bibfnamefont
  {A}~\bibnamefont {Datta}}, \bibinfo {author} {\bibfnamefont {N}~\bibnamefont
  {Thomas-Peter}}, \bibinfo {author} {\bibfnamefont {N~K}\ \bibnamefont
  {Langford}}, \bibinfo {author} {\bibfnamefont {D}~\bibnamefont {Kundys}},
  \bibinfo {author} {\bibfnamefont {J~C}\ \bibnamefont {Gates}}, \bibinfo
  {author} {\bibfnamefont {B~J}\ \bibnamefont {Smith}}, \bibinfo {author}
  {\bibfnamefont {P~G~R}\ \bibnamefont {Smith}}, \ and\ \bibinfo {author}
  {\bibfnamefont {I~A}\ \bibnamefont {Walmsley}},\ }\bibfield  {title}
  {\enquote {\bibinfo {title} {{Boson Sampling on a Photonic Chip}},}\
  }\href@noop {} {\bibfield  {journal} {\bibinfo  {journal} {Science}\ }\textbf
  {\bibinfo {volume} {339}},\ \bibinfo {pages} {798--801} (\bibinfo {year}
  {2013})}\BibitemShut {NoStop}%
\bibitem [{\citenamefont {Broome}\ \emph {et~al.}(2013)\citenamefont {Broome},
  \citenamefont {Fedrizzi}, \citenamefont {Rahimi-Keshari}, \citenamefont
  {Dove}, \citenamefont {Aaronson}, \citenamefont {Ralph},\ and\ \citenamefont
  {White}}]{Broome:2013bv}%
  \BibitemOpen
  \bibfield  {author} {\bibinfo {author} {\bibfnamefont {Matthew~A}\
  \bibnamefont {Broome}}, \bibinfo {author} {\bibfnamefont {Alessandro}\
  \bibnamefont {Fedrizzi}}, \bibinfo {author} {\bibfnamefont {Saleh}\
  \bibnamefont {Rahimi-Keshari}}, \bibinfo {author} {\bibfnamefont {Justin}\
  \bibnamefont {Dove}}, \bibinfo {author} {\bibfnamefont {Scott}\ \bibnamefont
  {Aaronson}}, \bibinfo {author} {\bibfnamefont {Timothy~C}\ \bibnamefont
  {Ralph}}, \ and\ \bibinfo {author} {\bibfnamefont {Andrew~G}\ \bibnamefont
  {White}},\ }\bibfield  {title} {\enquote {\bibinfo {title} {{Photonic Boson
  Sampling in a Tunable Circuit}},}\ }\href@noop {} {\bibfield  {journal}
  {\bibinfo  {journal} {Science}\ }\textbf {\bibinfo {volume} {339}},\ \bibinfo
  {pages} {794--798} (\bibinfo {year} {2013})}\BibitemShut {NoStop}%
\bibitem [{\citenamefont {Crespi}\ \emph {et~al.}(2013)\citenamefont {Crespi},
  \citenamefont {Osellame}, \citenamefont {Ramponi}, \citenamefont {Brod},
  \citenamefont {Galv{\~a}o}, \citenamefont {Spagnolo}, \citenamefont
  {Vitelli}, \citenamefont {Maiorino}, \citenamefont {Mataloni},\ and\
  \citenamefont {Sciarrino}}]{Crespi:2013fu}%
  \BibitemOpen
  \bibfield  {author} {\bibinfo {author} {\bibfnamefont {Andrea}\ \bibnamefont
  {Crespi}}, \bibinfo {author} {\bibfnamefont {Roberto}\ \bibnamefont
  {Osellame}}, \bibinfo {author} {\bibfnamefont {Roberta}\ \bibnamefont
  {Ramponi}}, \bibinfo {author} {\bibfnamefont {Daniel~J}\ \bibnamefont
  {Brod}}, \bibinfo {author} {\bibfnamefont {Ernesto~F}\ \bibnamefont
  {Galv{\~a}o}}, \bibinfo {author} {\bibfnamefont {Nicol{\`o}}\ \bibnamefont
  {Spagnolo}}, \bibinfo {author} {\bibfnamefont {Chiara}\ \bibnamefont
  {Vitelli}}, \bibinfo {author} {\bibfnamefont {Enrico}\ \bibnamefont
  {Maiorino}}, \bibinfo {author} {\bibfnamefont {Paolo}\ \bibnamefont
  {Mataloni}}, \ and\ \bibinfo {author} {\bibfnamefont {Fabio}\ \bibnamefont
  {Sciarrino}},\ }\bibfield  {title} {\enquote {\bibinfo {title} {{Integrated
  multimode interferometers with arbitrary designs for photonic boson
  sampling}},}\ }\href@noop {} {\bibfield  {journal} {\bibinfo  {journal}
  {Nature Photon}\ }\textbf {\bibinfo {volume} {7}},\ \bibinfo {pages}
  {545--549} (\bibinfo {year} {2013})}\BibitemShut {NoStop}%
\bibitem [{\citenamefont {Tillmann}\ \emph {et~al.}(2013)\citenamefont
  {Tillmann}, \citenamefont {Daki{\'c}}, \citenamefont {Heilmann},
  \citenamefont {Nolte}, \citenamefont {Szameit},\ and\ \citenamefont
  {Walther}}]{Tillmann:2013jv}%
  \BibitemOpen
  \bibfield  {author} {\bibinfo {author} {\bibfnamefont {Max}\ \bibnamefont
  {Tillmann}}, \bibinfo {author} {\bibfnamefont {Borivoje}\ \bibnamefont
  {Daki{\'c}}}, \bibinfo {author} {\bibfnamefont {Ren{\'e}}\ \bibnamefont
  {Heilmann}}, \bibinfo {author} {\bibfnamefont {Stefan}\ \bibnamefont
  {Nolte}}, \bibinfo {author} {\bibfnamefont {Alexander}\ \bibnamefont
  {Szameit}}, \ and\ \bibinfo {author} {\bibfnamefont {Philip}\ \bibnamefont
  {Walther}},\ }\bibfield  {title} {\enquote {\bibinfo {title} {{Experimental
  boson sampling}},}\ }\href@noop {} {\bibfield  {journal} {\bibinfo  {journal}
  {Nature Photon}\ }\textbf {\bibinfo {volume} {7}},\ \bibinfo {pages}
  {540--544} (\bibinfo {year} {2013})}\BibitemShut {NoStop}%
\bibitem [{\citenamefont {Li}\ \emph {et~al.}(2015)\citenamefont {Li},
  \citenamefont {Humphreys}, \citenamefont {Mendoza},\ and\ \citenamefont
  {Benjamin}}]{Li:2015ue}%
  \BibitemOpen
  \bibfield  {author} {\bibinfo {author} {\bibfnamefont {Ying}\ \bibnamefont
  {Li}}, \bibinfo {author} {\bibfnamefont {Peter~C}\ \bibnamefont {Humphreys}},
  \bibinfo {author} {\bibfnamefont {Gabriel~J}\ \bibnamefont {Mendoza}}, \ and\
  \bibinfo {author} {\bibfnamefont {Simon~C}\ \bibnamefont {Benjamin}},\
  }\bibfield  {title} {\enquote {\bibinfo {title} {{Resource costs for
  fault-tolerant linear optical quantum computing}},}\ }\href@noop {}
  {\bibfield  {journal} {\bibinfo  {journal} {arXiv}\ } (\bibinfo {year}
  {2015})},\ \Eprint {http://arxiv.org/abs/1504.02457} {1504.02457}
  \BibitemShut {NoStop}%
\bibitem [{\citenamefont {Kwiat}\ \emph {et~al.}(1995)\citenamefont {Kwiat},
  \citenamefont {Mattle}, \citenamefont {Weinfurter}, \citenamefont
  {Zeilinger}, \citenamefont {Sergienko},\ and\ \citenamefont
  {Shih}}]{Kwiat:1995ck}%
  \BibitemOpen
  \bibfield  {author} {\bibinfo {author} {\bibfnamefont {Paul~G.}\ \bibnamefont
  {Kwiat}}, \bibinfo {author} {\bibfnamefont {Klaus}\ \bibnamefont {Mattle}},
  \bibinfo {author} {\bibfnamefont {Harald}\ \bibnamefont {Weinfurter}},
  \bibinfo {author} {\bibfnamefont {Anton}\ \bibnamefont {Zeilinger}}, \bibinfo
  {author} {\bibfnamefont {Alexander~V.}\ \bibnamefont {Sergienko}}, \ and\
  \bibinfo {author} {\bibfnamefont {Yanhua}\ \bibnamefont {Shih}},\ }\bibfield
  {title} {\enquote {\bibinfo {title} {{New High-Intensity Source of
  Polarization-Entangled Photon Pairs}},}\ }\href@noop {} {\bibfield  {journal}
  {\bibinfo  {journal} {Phys. Rev. Lett.}\ }\textbf {\bibinfo {volume} {75}},\
  \bibinfo {pages} {4337--4341} (\bibinfo {year} {1995})}\BibitemShut {NoStop}%
\bibitem [{\citenamefont {Bechmann-Pasquinucci}\ and\ \citenamefont
  {Tittel}(2000)}]{BechmannPasquinucci:2000ug}%
  \BibitemOpen
  \bibfield  {author} {\bibinfo {author} {\bibfnamefont {H}~\bibnamefont
  {Bechmann-Pasquinucci}}\ and\ \bibinfo {author} {\bibfnamefont
  {W}~\bibnamefont {Tittel}},\ }\bibfield  {title} {\enquote {\bibinfo {title}
  {{Quantum cryptography using larger alphabets}},}\ }\href@noop {} {\bibfield
  {journal} {\bibinfo  {journal} {Phys. Rev. A}\ }\textbf {\bibinfo {volume}
  {61}},\ \bibinfo {pages} {062308} (\bibinfo {year} {2000})}\BibitemShut
  {NoStop}%
\bibitem [{\citenamefont {Cerf}\ \emph {et~al.}(2002)\citenamefont {Cerf},
  \citenamefont {Bourennane}, \citenamefont {Karlsson},\ and\ \citenamefont
  {Gisin}}]{Cerf:2002fp}%
  \BibitemOpen
  \bibfield  {author} {\bibinfo {author} {\bibfnamefont {Nicolas~J}\
  \bibnamefont {Cerf}}, \bibinfo {author} {\bibfnamefont {Mohamed}\
  \bibnamefont {Bourennane}}, \bibinfo {author} {\bibfnamefont {Anders}\
  \bibnamefont {Karlsson}}, \ and\ \bibinfo {author} {\bibfnamefont {Nicolas}\
  \bibnamefont {Gisin}},\ }\bibfield  {title} {\enquote {\bibinfo {title}
  {{Security of Quantum Key Distribution Using d-Level Systems}},}\ }\href@noop
  {} {\bibfield  {journal} {\bibinfo  {journal} {Phys. Rev. Lett.}\ }\textbf
  {\bibinfo {volume} {88}},\ \bibinfo {pages} {127902} (\bibinfo {year}
  {2002})}\BibitemShut {NoStop}%
\bibitem [{\citenamefont {Mair}\ \emph {et~al.}(2001)\citenamefont {Mair},
  \citenamefont {Vaziri}, \citenamefont {Weihs},\ and\ \citenamefont
  {Zeilinger}}]{Mair:2001fd}%
  \BibitemOpen
  \bibfield  {author} {\bibinfo {author} {\bibfnamefont {Alois}\ \bibnamefont
  {Mair}}, \bibinfo {author} {\bibfnamefont {Alipasha}\ \bibnamefont {Vaziri}},
  \bibinfo {author} {\bibfnamefont {Gregor}\ \bibnamefont {Weihs}}, \ and\
  \bibinfo {author} {\bibfnamefont {Anton}\ \bibnamefont {Zeilinger}},\
  }\bibfield  {title} {\enquote {\bibinfo {title} {{Entanglement of the orbital
  angular momentum states of photons}},}\ }\href@noop {} {\bibfield  {journal}
  {\bibinfo  {journal} {Nature}\ }\textbf {\bibinfo {volume} {412}},\ \bibinfo
  {pages} {313--316} (\bibinfo {year} {2001})}\BibitemShut {NoStop}%
\bibitem [{\citenamefont {Leach}\ \emph {et~al.}(2010)\citenamefont {Leach},
  \citenamefont {Jack}, \citenamefont {Romero}, \citenamefont {Jha},
  \citenamefont {Yao}, \citenamefont {Franke-Arnold}, \citenamefont {Ireland},
  \citenamefont {Boyd}, \citenamefont {Barnett},\ and\ \citenamefont
  {Padgett}}]{Leach:2010tt}%
  \BibitemOpen
  \bibfield  {author} {\bibinfo {author} {\bibfnamefont {Jonathan}\
  \bibnamefont {Leach}}, \bibinfo {author} {\bibfnamefont {Barry}\ \bibnamefont
  {Jack}}, \bibinfo {author} {\bibfnamefont {Jacqui}\ \bibnamefont {Romero}},
  \bibinfo {author} {\bibfnamefont {Anand~K}\ \bibnamefont {Jha}}, \bibinfo
  {author} {\bibfnamefont {Alison~M}\ \bibnamefont {Yao}}, \bibinfo {author}
  {\bibfnamefont {Sonja}\ \bibnamefont {Franke-Arnold}}, \bibinfo {author}
  {\bibfnamefont {David~G}\ \bibnamefont {Ireland}}, \bibinfo {author}
  {\bibfnamefont {Robert~W}\ \bibnamefont {Boyd}}, \bibinfo {author}
  {\bibfnamefont {Stephen~M}\ \bibnamefont {Barnett}}, \ and\ \bibinfo {author}
  {\bibfnamefont {Miles~J}\ \bibnamefont {Padgett}},\ }\bibfield  {title}
  {\enquote {\bibinfo {title} {{Quantum Correlations in Optical
  Angle{\textendash}Orbital Angular Momentum Variables}},}\ }\href@noop {}
  {\bibfield  {journal} {\bibinfo  {journal} {Science}\ }\textbf {\bibinfo
  {volume} {329}},\ \bibinfo {pages} {662--665} (\bibinfo {year}
  {2010})}\BibitemShut {NoStop}%
\bibitem [{\citenamefont {Dada}\ \emph {et~al.}(2011)\citenamefont {Dada},
  \citenamefont {Leach}, \citenamefont {Buller}, \citenamefont {Padgett},\ and\
  \citenamefont {Andersson}}]{Dada:2011do}%
  \BibitemOpen
  \bibfield  {author} {\bibinfo {author} {\bibfnamefont {Adetunmise~C}\
  \bibnamefont {Dada}}, \bibinfo {author} {\bibfnamefont {Jonathan}\
  \bibnamefont {Leach}}, \bibinfo {author} {\bibfnamefont {Gerald~S}\
  \bibnamefont {Buller}}, \bibinfo {author} {\bibfnamefont {Miles~J}\
  \bibnamefont {Padgett}}, \ and\ \bibinfo {author} {\bibfnamefont
  {E}~\bibnamefont {Andersson}},\ }\bibfield  {title} {\enquote {\bibinfo
  {title} {{Experimental high-dimensional two-photon entanglement and
  violations of generalized Bell inequalities}},}\ }\href@noop {} {\bibfield
  {journal} {\bibinfo  {journal} {Nature Phys.}\ }\textbf {\bibinfo {volume}
  {7}},\ \bibinfo {pages} {677--680} (\bibinfo {year} {2011})}\BibitemShut
  {NoStop}%
\bibitem [{\citenamefont {Berkhout}\ \emph {et~al.}(2010)\citenamefont
  {Berkhout}, \citenamefont {Lavery}, \citenamefont {Courtial}, \citenamefont
  {Beijersbergen},\ and\ \citenamefont {Padgett}}]{Berkhout:2010cb}%
  \BibitemOpen
  \bibfield  {author} {\bibinfo {author} {\bibfnamefont {Gregorius C~G}\
  \bibnamefont {Berkhout}}, \bibinfo {author} {\bibfnamefont {Martin P~J}\
  \bibnamefont {Lavery}}, \bibinfo {author} {\bibfnamefont {Johannes}\
  \bibnamefont {Courtial}}, \bibinfo {author} {\bibfnamefont {Marco~W}\
  \bibnamefont {Beijersbergen}}, \ and\ \bibinfo {author} {\bibfnamefont
  {Miles~J}\ \bibnamefont {Padgett}},\ }\bibfield  {title} {\enquote {\bibinfo
  {title} {{Efficient Sorting of Orbital Angular Momentum States of Light}},}\
  }\href@noop {} {\bibfield  {journal} {\bibinfo  {journal} {Phys. Rev. Lett.}\
  }\textbf {\bibinfo {volume} {105}},\ \bibinfo {pages} {153601} (\bibinfo
  {year} {2010})}\BibitemShut {NoStop}%
\bibitem [{\citenamefont {Gr{\"o}blacher}\ \emph {et~al.}(2006)\citenamefont
  {Gr{\"o}blacher}, \citenamefont {Jennewein}, \citenamefont {Vaziri},
  \citenamefont {Weihs},\ and\ \citenamefont {Zeilinger}}]{Groblacher:2006ec}%
  \BibitemOpen
  \bibfield  {author} {\bibinfo {author} {\bibfnamefont {Simon}\ \bibnamefont
  {Gr{\"o}blacher}}, \bibinfo {author} {\bibfnamefont {Thomas}\ \bibnamefont
  {Jennewein}}, \bibinfo {author} {\bibfnamefont {Alipasha}\ \bibnamefont
  {Vaziri}}, \bibinfo {author} {\bibfnamefont {Gregor}\ \bibnamefont {Weihs}},
  \ and\ \bibinfo {author} {\bibfnamefont {Anton}\ \bibnamefont {Zeilinger}},\
  }\bibfield  {title} {\enquote {\bibinfo {title} {{Experimental quantum
  cryptography with qutrits}},}\ }\href@noop {} {\bibfield  {journal} {\bibinfo
   {journal} {New J. Phys.}\ }\textbf {\bibinfo {volume} {8}},\ \bibinfo
  {pages} {75--75} (\bibinfo {year} {2006})}\BibitemShut {NoStop}%
\bibitem [{\citenamefont {Barreiro}\ \emph {et~al.}(2008)\citenamefont
  {Barreiro}, \citenamefont {Wei},\ and\ \citenamefont
  {Kwiat}}]{Barreiro:2008jl}%
  \BibitemOpen
  \bibfield  {author} {\bibinfo {author} {\bibfnamefont {Julio~T}\ \bibnamefont
  {Barreiro}}, \bibinfo {author} {\bibfnamefont {Tzu-Chieh}\ \bibnamefont
  {Wei}}, \ and\ \bibinfo {author} {\bibfnamefont {Paul~G}\ \bibnamefont
  {Kwiat}},\ }\bibfield  {title} {\enquote {\bibinfo {title} {{Beating the
  channel capacity limit for linear photonic superdense coding}},}\ }\href@noop
  {} {\bibfield  {journal} {\bibinfo  {journal} {Nature Phys.}\ }\textbf
  {\bibinfo {volume} {4}},\ \bibinfo {pages} {282--286} (\bibinfo {year}
  {2008})}\BibitemShut {NoStop}%
\bibitem [{\citenamefont {Leach}\ \emph {et~al.}(2012)\citenamefont {Leach},
  \citenamefont {Bolduc}, \citenamefont {Gauthier},\ and\ \citenamefont
  {Boyd}}]{Leach:2012gu}%
  \BibitemOpen
  \bibfield  {author} {\bibinfo {author} {\bibfnamefont {Jonathan}\
  \bibnamefont {Leach}}, \bibinfo {author} {\bibfnamefont {Eliot}\ \bibnamefont
  {Bolduc}}, \bibinfo {author} {\bibfnamefont {Daniel~J}\ \bibnamefont
  {Gauthier}}, \ and\ \bibinfo {author} {\bibfnamefont {Robert~W.}\
  \bibnamefont {Boyd}},\ }\bibfield  {title} {\enquote {\bibinfo {title}
  {{Secure information capacity of photons entangled in many dimensions}},}\
  }\href@noop {} {\bibfield  {journal} {\bibinfo  {journal} {Phys. Rev. A}\
  }\textbf {\bibinfo {volume} {85}},\ \bibinfo {pages} {060304} (\bibinfo
  {year} {2012})}\BibitemShut {NoStop}%
\bibitem [{\citenamefont {Titulaer}\ and\ \citenamefont
  {Glauber}(1966)}]{Titulaer:1966vb}%
  \BibitemOpen
  \bibfield  {author} {\bibinfo {author} {\bibfnamefont {U~M}\ \bibnamefont
  {Titulaer}}\ and\ \bibinfo {author} {\bibfnamefont {R~J}\ \bibnamefont
  {Glauber}},\ }\bibfield  {title} {\enquote {\bibinfo {title} {{Density
  operators for coherent fields}},}\ }\href@noop {} {\bibfield  {journal}
  {\bibinfo  {journal} {Phys. Rev.}\ }\textbf {\bibinfo {volume} {145}},\
  \bibinfo {pages} {1041} (\bibinfo {year} {1966})}\BibitemShut {NoStop}%
\bibitem [{\citenamefont {Sohler}\ \emph {et~al.}(2013)\citenamefont {Sohler},
  \citenamefont {Ostrowsky}, \citenamefont {Alibart},\ and\ \citenamefont
  {Tanzilli}}]{Roslund:2013cb}%
  \BibitemOpen
  \bibfield  {author} {\bibinfo {author} {\bibfnamefont {W}~\bibnamefont
  {Sohler}}, \bibinfo {author} {\bibfnamefont {D}~\bibnamefont {Ostrowsky}},
  \bibinfo {author} {\bibfnamefont {O}~\bibnamefont {Alibart}}, \ and\ \bibinfo
  {author} {\bibfnamefont {S}~\bibnamefont {Tanzilli}},\ }\bibfield  {title}
  {\enquote {\bibinfo {title} {{Wavelength-multiplexed quantum networks with
  ultrafast frequency combs}},}\ }\href@noop {} {\bibfield  {journal} {\bibinfo
   {journal} {Nature Photon}\ }\textbf {\bibinfo {volume} {8}},\ \bibinfo
  {pages} {109--112} (\bibinfo {year} {2013})}\BibitemShut {NoStop}%
\bibitem [{\citenamefont {Averchenko}\ \emph {et~al.}(2014)\citenamefont
  {Averchenko}, \citenamefont {Thiel},\ and\ \citenamefont
  {Treps}}]{Averchenko:2014dn}%
  \BibitemOpen
  \bibfield  {author} {\bibinfo {author} {\bibfnamefont {V~A}\ \bibnamefont
  {Averchenko}}, \bibinfo {author} {\bibfnamefont {V}~\bibnamefont {Thiel}}, \
  and\ \bibinfo {author} {\bibfnamefont {N}~\bibnamefont {Treps}},\ }\bibfield
  {title} {\enquote {\bibinfo {title} {{Non-linear photon subtraction from a
  multimode quantum field}},}\ }\href@noop {} {\bibfield  {journal} {\bibinfo
  {journal} {Phys. Rev. A}\ }\textbf {\bibinfo {volume} {89}},\ \bibinfo
  {pages} {063808} (\bibinfo {year} {2014})}\BibitemShut {NoStop}%
\bibitem [{\citenamefont {Smith}\ and\ \citenamefont
  {Raymer}(2007)}]{Smith:2007dq}%
  \BibitemOpen
  \bibfield  {author} {\bibinfo {author} {\bibfnamefont {Brian~J}\ \bibnamefont
  {Smith}}\ and\ \bibinfo {author} {\bibfnamefont {M~G}\ \bibnamefont
  {Raymer}},\ }\bibfield  {title} {\enquote {\bibinfo {title} {{Photon wave
  functions, wave-packet quantization of light, and coherence theory}},}\
  }\href@noop {} {\bibfield  {journal} {\bibinfo  {journal} {New J. Phys.}\
  }\textbf {\bibinfo {volume} {9}},\ \bibinfo {pages} {414--414} (\bibinfo
  {year} {2007})}\BibitemShut {NoStop}%
\bibitem [{\citenamefont {Schwinger}(1960)}]{Schwinger:1960wq}%
  \BibitemOpen
  \bibfield  {author} {\bibinfo {author} {\bibfnamefont {Julian}\ \bibnamefont
  {Schwinger}},\ }\bibfield  {title} {\enquote {\bibinfo {title} {{Unitary
  Operator Bases}},}\ }\href@noop {} {\bibfield  {journal} {\bibinfo  {journal}
  {Proc. Natl. Acad. Sci. U.S.A.}\ }\textbf {\bibinfo {volume} {46}},\ \bibinfo
  {pages} {570} (\bibinfo {year} {1960})}\BibitemShut {NoStop}%
\bibitem [{\citenamefont {Bennett}\ and\ \citenamefont
  {Brassard}(1984)}]{Bennett:1984wv}%
  \BibitemOpen
  \bibfield  {author} {\bibinfo {author} {\bibfnamefont {Charles~H}\
  \bibnamefont {Bennett}}\ and\ \bibinfo {author} {\bibfnamefont {Gilles}\
  \bibnamefont {Brassard}},\ }\bibfield  {title} {\enquote {\bibinfo {title}
  {{Quantum cryptography: Public key distribution and coin tossing}},}\ }in\
  \href@noop {} {\emph {\bibinfo {booktitle} {Proceedings of IEEE International
  Conference on Computers, Systems and Signal Processing, Bangalore, India}}}\
  (\bibinfo {year} {1984})\ pp.\ \bibinfo {pages} {175--179}\BibitemShut
  {NoStop}%
\bibitem [{\citenamefont {Smithey}\ \emph {et~al.}(1993)\citenamefont
  {Smithey}, \citenamefont {Beck}, \citenamefont {Raymer},\ and\ \citenamefont
  {Faridani}}]{Smithey:1993er}%
  \BibitemOpen
  \bibfield  {author} {\bibinfo {author} {\bibfnamefont {D.~T.}\ \bibnamefont
  {Smithey}}, \bibinfo {author} {\bibfnamefont {M}~\bibnamefont {Beck}},
  \bibinfo {author} {\bibfnamefont {Michael~G.}\ \bibnamefont {Raymer}}, \ and\
  \bibinfo {author} {\bibfnamefont {A}~\bibnamefont {Faridani}},\ }\bibfield
  {title} {\enquote {\bibinfo {title} {{Measurement of the Wigner distribution
  and the density matrix of a light mode using optical homodyne tomography:
  Application to squeezed states and the vacuum}},}\ }\href@noop {} {\bibfield
  {journal} {\bibinfo  {journal} {Phys. Rev. Lett.}\ }\textbf {\bibinfo
  {volume} {70}},\ \bibinfo {pages} {1244--1247} (\bibinfo {year}
  {1993})}\BibitemShut {NoStop}%
\bibitem [{\citenamefont {Eckstein}\ \emph
  {et~al.}(2011{\natexlab{a}})\citenamefont {Eckstein}, \citenamefont
  {Brecht},\ and\ \citenamefont {Silberhorn}}]{EcksteinA:2011vg}%
  \BibitemOpen
  \bibfield  {author} {\bibinfo {author} {\bibfnamefont {Andreas}\ \bibnamefont
  {Eckstein}}, \bibinfo {author} {\bibfnamefont {Benjamin}\ \bibnamefont
  {Brecht}}, \ and\ \bibinfo {author} {\bibfnamefont {Christine}\ \bibnamefont
  {Silberhorn}},\ }\bibfield  {title} {\enquote {\bibinfo {title} {{A quantum
  pulse gate based on spectrally engineered sum frequency generation}},}\
  }\href@noop {} {\bibfield  {journal} {\bibinfo  {journal} {Opt. Express}\
  }\textbf {\bibinfo {volume} {19}},\ \bibinfo {pages} {13770} (\bibinfo {year}
  {2011}{\natexlab{a}})}\BibitemShut {NoStop}%
\bibitem [{\citenamefont {Brecht}\ \emph {et~al.}(2014)\citenamefont {Brecht},
  \citenamefont {Eckstein}, \citenamefont {Ricken}, \citenamefont {Quiring},
  \citenamefont {Suche}, \citenamefont {Sansoni},\ and\ \citenamefont
  {Silberhorn}}]{Brecht:2014eg}%
  \BibitemOpen
  \bibfield  {author} {\bibinfo {author} {\bibfnamefont {Benjamin}\
  \bibnamefont {Brecht}}, \bibinfo {author} {\bibfnamefont {Andreas}\
  \bibnamefont {Eckstein}}, \bibinfo {author} {\bibfnamefont {Raimund}\
  \bibnamefont {Ricken}}, \bibinfo {author} {\bibfnamefont {Viktor}\
  \bibnamefont {Quiring}}, \bibinfo {author} {\bibfnamefont {Hubertus}\
  \bibnamefont {Suche}}, \bibinfo {author} {\bibfnamefont {Linda}\ \bibnamefont
  {Sansoni}}, \ and\ \bibinfo {author} {\bibfnamefont {Christine}\ \bibnamefont
  {Silberhorn}},\ }\bibfield  {title} {\enquote {\bibinfo {title}
  {{Demonstration of coherent time-frequency Schmidt mode selection using
  dispersion-engineered frequency conversion}},}\ }\href@noop {} {\bibfield
  {journal} {\bibinfo  {journal} {Phys. Rev. A}\ }\textbf {\bibinfo {volume}
  {90}},\ \bibinfo {pages} {030302(R)} (\bibinfo {year} {2014})}\BibitemShut
  {NoStop}%
\bibitem [{\citenamefont {Reddy}\ \emph {et~al.}(2014)\citenamefont {Reddy},
  \citenamefont {Raymer},\ and\ \citenamefont {McKinstrie}}]{Reddy:2014bt}%
  \BibitemOpen
  \bibfield  {author} {\bibinfo {author} {\bibfnamefont {D~V}\ \bibnamefont
  {Reddy}}, \bibinfo {author} {\bibfnamefont {M~G}\ \bibnamefont {Raymer}}, \
  and\ \bibinfo {author} {\bibfnamefont {C~J}\ \bibnamefont {McKinstrie}},\
  }\bibfield  {title} {\enquote {\bibinfo {title} {{Efficient sorting of
  quantum-optical wave packets by temporal-mode interferometry}},}\ }\href@noop
  {} {\bibfield  {journal} {\bibinfo  {journal} {Opt. Lett.}\ }\textbf
  {\bibinfo {volume} {39}},\ \bibinfo {pages} {2924--2927} (\bibinfo {year}
  {2014})}\BibitemShut {NoStop}%
\bibitem [{\citenamefont {Reddy}\ \emph {et~al.}(2015)\citenamefont {Reddy},
  \citenamefont {Raymer},\ and\ \citenamefont {McKinstrie}}]{Reddy:2015jb}%
  \BibitemOpen
  \bibfield  {author} {\bibinfo {author} {\bibfnamefont {D~V}\ \bibnamefont
  {Reddy}}, \bibinfo {author} {\bibfnamefont {M~G}\ \bibnamefont {Raymer}}, \
  and\ \bibinfo {author} {\bibfnamefont {C~J}\ \bibnamefont {McKinstrie}},\
  }\bibfield  {title} {\enquote {\bibinfo {title} {{Sorting photon wave packets
  using temporal-mode interferometry based on multiple-stage quantum frequency
  conversion}},}\ }\href@noop {} {\bibfield  {journal} {\bibinfo  {journal}
  {Phys. Rev. A}\ }\textbf {\bibinfo {volume} {91}},\ \bibinfo {pages} {012323}
  (\bibinfo {year} {2015})}\BibitemShut {NoStop}%
\bibitem [{\citenamefont {Law}\ \emph {et~al.}(2000)\citenamefont {Law},
  \citenamefont {Walmsley},\ and\ \citenamefont {Eberly}}]{Law:2000wd}%
  \BibitemOpen
  \bibfield  {author} {\bibinfo {author} {\bibfnamefont {C~K}\ \bibnamefont
  {Law}}, \bibinfo {author} {\bibfnamefont {I~A}\ \bibnamefont {Walmsley}}, \
  and\ \bibinfo {author} {\bibfnamefont {J~H}\ \bibnamefont {Eberly}},\
  }\bibfield  {title} {\enquote {\bibinfo {title} {{Continuous Frequency
  Entanglement: Effective Finite Hilbert Space and Entropy Control}},}\
  }\href@noop {} {\bibfield  {journal} {\bibinfo  {journal} {Phys. Rev. Lett.}\
  }\textbf {\bibinfo {volume} {84}},\ \bibinfo {pages} {5304--5307} (\bibinfo
  {year} {2000})}\BibitemShut {NoStop}%
\bibitem [{\citenamefont {Grice}\ and\ \citenamefont
  {Walmsley}(1997)}]{Grice:1997tk}%
  \BibitemOpen
  \bibfield  {author} {\bibinfo {author} {\bibfnamefont {W~P}\ \bibnamefont
  {Grice}}\ and\ \bibinfo {author} {\bibfnamefont {I~A}\ \bibnamefont
  {Walmsley}},\ }\bibfield  {title} {\enquote {\bibinfo {title} {{Spectral
  information and distinguishability in type-II down-conversion with a
  broadband pump}},}\ }\href@noop {} {\bibfield  {journal} {\bibinfo  {journal}
  {Phys. Rev. A}\ }\textbf {\bibinfo {volume} {56}},\ \bibinfo {pages}
  {1627--1634} (\bibinfo {year} {1997})}\BibitemShut {NoStop}%
\bibitem [{\citenamefont {Keller}\ and\ \citenamefont
  {Rubin}(1997)}]{Keller:1997hj}%
  \BibitemOpen
  \bibfield  {author} {\bibinfo {author} {\bibfnamefont {Timothy~E}\
  \bibnamefont {Keller}}\ and\ \bibinfo {author} {\bibfnamefont {Morton~H}\
  \bibnamefont {Rubin}},\ }\bibfield  {title} {\enquote {\bibinfo {title}
  {{Theory of two-photon entanglement for spontaneous parametric
  down-conversion driven by a narrow pump pulse}},}\ }\href@noop {} {\bibfield
  {journal} {\bibinfo  {journal} {Phys. Rev. A}\ }\textbf {\bibinfo {volume}
  {56}},\ \bibinfo {pages} {1534--1541} (\bibinfo {year} {1997})}\BibitemShut
  {NoStop}%
\bibitem [{\citenamefont {U'Ren}\ \emph {et~al.}(2005)\citenamefont {U'Ren},
  \citenamefont {Silberhorn}, \citenamefont {Banaszek}, \citenamefont
  {Walmsley}, \citenamefont {Erdmann}, \citenamefont {Grice},\ and\
  \citenamefont {Raymer}}]{URen:2005wb}%
  \BibitemOpen
  \bibfield  {author} {\bibinfo {author} {\bibfnamefont {A~B}\ \bibnamefont
  {U'Ren}}, \bibinfo {author} {\bibfnamefont {C}~\bibnamefont {Silberhorn}},
  \bibinfo {author} {\bibfnamefont {Konrad}\ \bibnamefont {Banaszek}}, \bibinfo
  {author} {\bibfnamefont {I~A}\ \bibnamefont {Walmsley}}, \bibinfo {author}
  {\bibfnamefont {R.}~\bibnamefont {Erdmann}}, \bibinfo {author} {\bibfnamefont
  {W~P}\ \bibnamefont {Grice}}, \ and\ \bibinfo {author} {\bibfnamefont {M~G}\
  \bibnamefont {Raymer}},\ }\bibfield  {title} {\enquote {\bibinfo {title}
  {{Generation of Pure-State Single-Photon Wavepackets by Conditional
  Preparation Based on Spontaneous Parametric Downconversion}},}\ }\href@noop
  {} {\bibfield  {journal} {\bibinfo  {journal} {Laser Phys.}\ }\textbf
  {\bibinfo {volume} {15}},\ \bibinfo {pages} {146--161} (\bibinfo {year}
  {2005})}\BibitemShut {NoStop}%
\bibitem [{\citenamefont {Mosley}\ \emph {et~al.}(2008)\citenamefont {Mosley},
  \citenamefont {Lundeen}, \citenamefont {Smith}, \citenamefont {Wasylczyk},
  \citenamefont {U'Ren}, \citenamefont {Silberhorn},\ and\ \citenamefont
  {Walmsley}}]{Mosley:2008ir}%
  \BibitemOpen
  \bibfield  {author} {\bibinfo {author} {\bibfnamefont {P~J}\ \bibnamefont
  {Mosley}}, \bibinfo {author} {\bibfnamefont {J~S}\ \bibnamefont {Lundeen}},
  \bibinfo {author} {\bibfnamefont {B~J}\ \bibnamefont {Smith}}, \bibinfo
  {author} {\bibfnamefont {P}~\bibnamefont {Wasylczyk}}, \bibinfo {author}
  {\bibfnamefont {A~B}\ \bibnamefont {U'Ren}}, \bibinfo {author} {\bibfnamefont
  {Christine}\ \bibnamefont {Silberhorn}}, \ and\ \bibinfo {author}
  {\bibfnamefont {Ian~A}\ \bibnamefont {Walmsley}},\ }\bibfield  {title}
  {\enquote {\bibinfo {title} {{Heralded generation of ultrafast single photons
  in pure quantum states}},}\ }\href@noop {} {\bibfield  {journal} {\bibinfo
  {journal} {Phys. Rev. Lett.}\ }\textbf {\bibinfo {volume} {100}},\ \bibinfo
  {pages} {133601} (\bibinfo {year} {2008})}\BibitemShut {NoStop}%
\bibitem [{\citenamefont {Kuzucu}\ \emph {et~al.}(2008)\citenamefont {Kuzucu},
  \citenamefont {Wong}, \citenamefont {Kurimura},\ and\ \citenamefont
  {Tovstonog}}]{Kuzucu:2008gv}%
  \BibitemOpen
  \bibfield  {author} {\bibinfo {author} {\bibfnamefont {Onur}\ \bibnamefont
  {Kuzucu}}, \bibinfo {author} {\bibfnamefont {Franco N~C}\ \bibnamefont
  {Wong}}, \bibinfo {author} {\bibfnamefont {Sunao}\ \bibnamefont {Kurimura}},
  \ and\ \bibinfo {author} {\bibfnamefont {Sergey}\ \bibnamefont {Tovstonog}},\
  }\bibfield  {title} {\enquote {\bibinfo {title} {{Joint Temporal Density
  Measurements for Two-Photon State Characterization}},}\ }\href@noop {}
  {\bibfield  {journal} {\bibinfo  {journal} {Phys. Rev. Lett.}\ }\textbf
  {\bibinfo {volume} {101}},\ \bibinfo {pages} {153602} (\bibinfo {year}
  {2008})}\BibitemShut {NoStop}%
\bibitem [{\citenamefont {Shi}\ \emph {et~al.}(2008)\citenamefont {Shi},
  \citenamefont {Valencia}, \citenamefont {Hendrych},\ and\ \citenamefont
  {Torres}}]{Shi:2008tl}%
  \BibitemOpen
  \bibfield  {author} {\bibinfo {author} {\bibfnamefont {X}~\bibnamefont
  {Shi}}, \bibinfo {author} {\bibfnamefont {A}~\bibnamefont {Valencia}},
  \bibinfo {author} {\bibfnamefont {M}~\bibnamefont {Hendrych}}, \ and\
  \bibinfo {author} {\bibfnamefont {J~P}\ \bibnamefont {Torres}},\ }\bibfield
  {title} {\enquote {\bibinfo {title} {{Generation of indistinguishable and
  pure heralded single photons with tunable bandwidth}},}\ }\href@noop {}
  {\bibfield  {journal} {\bibinfo  {journal} {Opt. Lett.}\ }\textbf {\bibinfo
  {volume} {33}},\ \bibinfo {pages} {875--877} (\bibinfo {year}
  {2008})}\BibitemShut {NoStop}%
\bibitem [{\citenamefont {Eckstein}\ \emph
  {et~al.}(2011{\natexlab{b}})\citenamefont {Eckstein}, \citenamefont {Christ},
  \citenamefont {Mosley},\ and\ \citenamefont {Silberhorn}}]{Eckstein:2011wl}%
  \BibitemOpen
  \bibfield  {author} {\bibinfo {author} {\bibfnamefont {Andreas}\ \bibnamefont
  {Eckstein}}, \bibinfo {author} {\bibfnamefont {Andreas}\ \bibnamefont
  {Christ}}, \bibinfo {author} {\bibfnamefont {Peter~J}\ \bibnamefont
  {Mosley}}, \ and\ \bibinfo {author} {\bibfnamefont {Christine}\ \bibnamefont
  {Silberhorn}},\ }\bibfield  {title} {\enquote {\bibinfo {title} {{Highly
  Efficient Single-Pass Source of Pulsed Single-Mode Twin Beams of Light}},}\
  }\href@noop {} {\bibfield  {journal} {\bibinfo  {journal} {Phys. Rev. Lett.}\
  }\textbf {\bibinfo {volume} {106}},\ \bibinfo {pages} {013603} (\bibinfo
  {year} {2011}{\natexlab{b}})}\BibitemShut {NoStop}%
\bibitem [{\citenamefont {Harder}\ \emph {et~al.}(2013)\citenamefont {Harder},
  \citenamefont {Ansari}, \citenamefont {Brecht}, \citenamefont {Dirmeier},
  \citenamefont {Marquardt},\ and\ \citenamefont {Silberhorn}}]{Harder:2013hk}%
  \BibitemOpen
  \bibfield  {author} {\bibinfo {author} {\bibfnamefont {Georg}\ \bibnamefont
  {Harder}}, \bibinfo {author} {\bibfnamefont {Vahid}\ \bibnamefont {Ansari}},
  \bibinfo {author} {\bibfnamefont {Benjamin}\ \bibnamefont {Brecht}}, \bibinfo
  {author} {\bibfnamefont {Thomas}\ \bibnamefont {Dirmeier}}, \bibinfo {author}
  {\bibfnamefont {Christoph}\ \bibnamefont {Marquardt}}, \ and\ \bibinfo
  {author} {\bibfnamefont {Christine}\ \bibnamefont {Silberhorn}},\ }\bibfield
  {title} {\enquote {\bibinfo {title} {{An optimized photon pair source for
  quantum circuits}},}\ }\href@noop {} {\bibfield  {journal} {\bibinfo
  {journal} {Opt. Express}\ }\textbf {\bibinfo {volume} {21}},\ \bibinfo
  {pages} {13975--13985} (\bibinfo {year} {2013})}\BibitemShut {NoStop}%
\bibitem [{\citenamefont {Brecht}\ \emph {et~al.}(2011)\citenamefont {Brecht},
  \citenamefont {Eckstein}, \citenamefont {Christ}, \citenamefont {Suche},\
  and\ \citenamefont {Silberhorn}}]{Brecht:2011hz}%
  \BibitemOpen
  \bibfield  {author} {\bibinfo {author} {\bibfnamefont {Benjamin}\
  \bibnamefont {Brecht}}, \bibinfo {author} {\bibfnamefont {Andreas}\
  \bibnamefont {Eckstein}}, \bibinfo {author} {\bibfnamefont {Andreas}\
  \bibnamefont {Christ}}, \bibinfo {author} {\bibfnamefont {Hubertus}\
  \bibnamefont {Suche}}, \ and\ \bibinfo {author} {\bibfnamefont {Christine}\
  \bibnamefont {Silberhorn}},\ }\bibfield  {title} {\enquote {\bibinfo {title}
  {{From quantum pulse gate to quantum pulse shaper{\textemdash}engineered
  frequency conversion in nonlinear optical waveguides}},}\ }\href@noop {}
  {\bibfield  {journal} {\bibinfo  {journal} {New J. Phys.}\ }\textbf {\bibinfo
  {volume} {13}},\ \bibinfo {pages} {065029} (\bibinfo {year}
  {2011})}\BibitemShut {NoStop}%
\bibitem [{\citenamefont {M{\'e}chin}\ \emph {et~al.}(2006)\citenamefont
  {M{\'e}chin}, \citenamefont {Provo}, \citenamefont {Harvey},\ and\
  \citenamefont {McKinstrie}}]{Mechin:2006du}%
  \BibitemOpen
  \bibfield  {author} {\bibinfo {author} {\bibfnamefont {D}~\bibnamefont
  {M{\'e}chin}}, \bibinfo {author} {\bibfnamefont {R}~\bibnamefont {Provo}},
  \bibinfo {author} {\bibfnamefont {J~D}\ \bibnamefont {Harvey}}, \ and\
  \bibinfo {author} {\bibfnamefont {C~J}\ \bibnamefont {McKinstrie}},\
  }\bibfield  {title} {\enquote {\bibinfo {title} {{180-nm wavelength
  conversion based on Bragg scattering in an optical fiber}},}\ }\href@noop {}
  {\bibfield  {journal} {\bibinfo  {journal} {Opt. Express}\ }\textbf {\bibinfo
  {volume} {14}},\ \bibinfo {pages} {8995--8999} (\bibinfo {year}
  {2006})}\BibitemShut {NoStop}%
\bibitem [{\citenamefont {McGuinness}\ \emph {et~al.}(2011)\citenamefont
  {McGuinness}, \citenamefont {Raymer}, \citenamefont {McKinstrie},\ and\
  \citenamefont {Radic}}]{McGuinness:2011fv}%
  \BibitemOpen
  \bibfield  {author} {\bibinfo {author} {\bibfnamefont {H~J}\ \bibnamefont
  {McGuinness}}, \bibinfo {author} {\bibfnamefont {M~G}\ \bibnamefont
  {Raymer}}, \bibinfo {author} {\bibfnamefont {C~J}\ \bibnamefont
  {McKinstrie}}, \ and\ \bibinfo {author} {\bibfnamefont {S}~\bibnamefont
  {Radic}},\ }\bibfield  {title} {\enquote {\bibinfo {title} {{Wavelength
  Translation Across 210 nm in the Visible Using Vector Bragg Scattering in a
  Birefringent Photonic Crystal Fiber}},}\ }\href@noop {} {\bibfield  {journal}
  {\bibinfo  {journal} {IEEE Photon. Technol. Lett.}\ }\textbf {\bibinfo
  {volume} {23}},\ \bibinfo {pages} {109--111} (\bibinfo {year}
  {2011})}\BibitemShut {NoStop}%
\bibitem [{\citenamefont {McGuinness}\ \emph {et~al.}(2010)\citenamefont
  {McGuinness}, \citenamefont {Raymer}, \citenamefont {McKinstrie},\ and\
  \citenamefont {Radic}}]{McGuinness:2010ja}%
  \BibitemOpen
  \bibfield  {author} {\bibinfo {author} {\bibfnamefont {H~J}\ \bibnamefont
  {McGuinness}}, \bibinfo {author} {\bibfnamefont {M~G}\ \bibnamefont
  {Raymer}}, \bibinfo {author} {\bibfnamefont {C~J}\ \bibnamefont
  {McKinstrie}}, \ and\ \bibinfo {author} {\bibfnamefont {S}~\bibnamefont
  {Radic}},\ }\bibfield  {title} {\enquote {\bibinfo {title} {{Quantum
  Frequency Translation of Single-Photon States in a Photonic Crystal
  Fiber}},}\ }\href@noop {} {\bibfield  {journal} {\bibinfo  {journal} {Phys.
  Rev. Lett.}\ }\textbf {\bibinfo {volume} {105}},\ \bibinfo {pages} {093604}
  (\bibinfo {year} {2010})}\BibitemShut {NoStop}%
\bibitem [{\citenamefont {Raymer}\ \emph {et~al.}(2010)\citenamefont {Raymer},
  \citenamefont {van Enk}, \citenamefont {McKinstrie},\ and\ \citenamefont
  {McGuinness}}]{Raymer:2010df}%
  \BibitemOpen
  \bibfield  {author} {\bibinfo {author} {\bibfnamefont {M~G}\ \bibnamefont
  {Raymer}}, \bibinfo {author} {\bibfnamefont {S~J}\ \bibnamefont {van Enk}},
  \bibinfo {author} {\bibfnamefont {C~J}\ \bibnamefont {McKinstrie}}, \ and\
  \bibinfo {author} {\bibfnamefont {H~J}\ \bibnamefont {McGuinness}},\
  }\bibfield  {title} {\enquote {\bibinfo {title} {{Interference of two photons
  of different color}},}\ }\href@noop {} {\bibfield  {journal} {\bibinfo
  {journal} {Optics Communications}\ }\textbf {\bibinfo {volume} {283}},\
  \bibinfo {pages} {747--752} (\bibinfo {year} {2010})}\BibitemShut {NoStop}%
\bibitem [{\citenamefont {Reddy}\ \emph {et~al.}(2013)\citenamefont {Reddy},
  \citenamefont {Raymer}, \citenamefont {McKinstrie}, \citenamefont {Mejling},\
  and\ \citenamefont {Rottwitt}}]{Reddy:2013ip}%
  \BibitemOpen
  \bibfield  {author} {\bibinfo {author} {\bibfnamefont {D~V}\ \bibnamefont
  {Reddy}}, \bibinfo {author} {\bibfnamefont {M~G}\ \bibnamefont {Raymer}},
  \bibinfo {author} {\bibfnamefont {C~J}\ \bibnamefont {McKinstrie}}, \bibinfo
  {author} {\bibfnamefont {L.}~\bibnamefont {Mejling}}, \ and\ \bibinfo
  {author} {\bibfnamefont {K.}~\bibnamefont {Rottwitt}},\ }\bibfield  {title}
  {\enquote {\bibinfo {title} {{Temporal mode selectivity by frequency
  conversion in second-order nonlinear optical waveguides}},}\ }\href@noop {}
  {\bibfield  {journal} {\bibinfo  {journal} {Opt. Express}\ }\textbf {\bibinfo
  {volume} {21}},\ \bibinfo {pages} {13840--13863} (\bibinfo {year}
  {2013})}\BibitemShut {NoStop}%
\bibitem [{\citenamefont {Christ}\ \emph {et~al.}(2013)\citenamefont {Christ},
  \citenamefont {Brecht}, \citenamefont {Mauerer},\ and\ \citenamefont
  {Silberhorn}}]{Christ:2013fg}%
  \BibitemOpen
  \bibfield  {author} {\bibinfo {author} {\bibfnamefont {Andreas}\ \bibnamefont
  {Christ}}, \bibinfo {author} {\bibfnamefont {Benjamin}\ \bibnamefont
  {Brecht}}, \bibinfo {author} {\bibfnamefont {Wolfgang}\ \bibnamefont
  {Mauerer}}, \ and\ \bibinfo {author} {\bibfnamefont {Christine}\ \bibnamefont
  {Silberhorn}},\ }\bibfield  {title} {\enquote {\bibinfo {title} {{Theory of
  quantum frequency conversion and type-II parametric down-conversion in the
  high-gain regime}},}\ }\href@noop {} {\bibfield  {journal} {\bibinfo
  {journal} {New J. Phys.}\ }\textbf {\bibinfo {volume} {15}},\ \bibinfo
  {pages} {053038} (\bibinfo {year} {2013})}\BibitemShut {NoStop}%
\bibitem [{\citenamefont {Quesada}\ and\ \citenamefont
  {Sipe}(2014)}]{Quesada:2014gn}%
  \BibitemOpen
  \bibfield  {author} {\bibinfo {author} {\bibfnamefont {Nicol{\'a}s}\
  \bibnamefont {Quesada}}\ and\ \bibinfo {author} {\bibfnamefont {J~E}\
  \bibnamefont {Sipe}},\ }\bibfield  {title} {\enquote {\bibinfo {title}
  {{Effects of time ordering in quantum nonlinear optics}},}\ }\href@noop {}
  {\bibfield  {journal} {\bibinfo  {journal} {Phys. Rev. A}\ }\textbf {\bibinfo
  {volume} {90}},\ \bibinfo {pages} {063840} (\bibinfo {year}
  {2014})}\BibitemShut {NoStop}%
\bibitem [{\citenamefont {Huang}\ and\ \citenamefont
  {Kumar}(2013)}]{Huang:2013tu}%
  \BibitemOpen
  \bibfield  {author} {\bibinfo {author} {\bibfnamefont {Y-P}\ \bibnamefont
  {Huang}}\ and\ \bibinfo {author} {\bibfnamefont {P}~\bibnamefont {Kumar}},\
  }\bibfield  {title} {\enquote {\bibinfo {title} {{Mode-resolved photon
  counting via cascaded quantum frequency conversion}},}\ }\href@noop {}
  {\bibfield  {journal} {\bibinfo  {journal} {Opt. Lett.}\ }\textbf {\bibinfo
  {volume} {38}},\ \bibinfo {pages} {468--470} (\bibinfo {year}
  {2013})}\BibitemShut {NoStop}%
\bibitem [{\citenamefont {Kowligy}\ \emph {et~al.}(2014)\citenamefont
  {Kowligy}, \citenamefont {Manurkar}, \citenamefont {Corzo}, \citenamefont
  {Velev}, \citenamefont {Silver}, \citenamefont {Scott}, \citenamefont {Yoo},
  \citenamefont {Kumar}, \citenamefont {Kanter},\ and\ \citenamefont
  {Huang}}]{Kowligy:2014ga}%
  \BibitemOpen
  \bibfield  {author} {\bibinfo {author} {\bibfnamefont {Abijith~S}\
  \bibnamefont {Kowligy}}, \bibinfo {author} {\bibfnamefont {Paritosh}\
  \bibnamefont {Manurkar}}, \bibinfo {author} {\bibfnamefont {Neil~V}\
  \bibnamefont {Corzo}}, \bibinfo {author} {\bibfnamefont {Vesselin~G}\
  \bibnamefont {Velev}}, \bibinfo {author} {\bibfnamefont {Michael}\
  \bibnamefont {Silver}}, \bibinfo {author} {\bibfnamefont {Ryan~P}\
  \bibnamefont {Scott}}, \bibinfo {author} {\bibfnamefont {S~J~B}\ \bibnamefont
  {Yoo}}, \bibinfo {author} {\bibfnamefont {Prem}\ \bibnamefont {Kumar}},
  \bibinfo {author} {\bibfnamefont {Gregory~S}\ \bibnamefont {Kanter}}, \ and\
  \bibinfo {author} {\bibfnamefont {Yu-Ping}\ \bibnamefont {Huang}},\
  }\bibfield  {title} {\enquote {\bibinfo {title} {{Quantum optical arbitrary
  waveform manipulation and measurement in real time}},}\ }\href@noop {}
  {\bibfield  {journal} {\bibinfo  {journal} {Opt. Express}\ }\textbf {\bibinfo
  {volume} {22}},\ \bibinfo {pages} {27942--27957} (\bibinfo {year}
  {2014})}\BibitemShut {NoStop}%
\bibitem [{\citenamefont {Donohue}\ \emph {et~al.}(2015)\citenamefont
  {Donohue}, \citenamefont {Mazurek},\ and\ \citenamefont
  {Resch}}]{Donohue:2015cd}%
  \BibitemOpen
  \bibfield  {author} {\bibinfo {author} {\bibfnamefont {John~M}\ \bibnamefont
  {Donohue}}, \bibinfo {author} {\bibfnamefont {Michael~D}\ \bibnamefont
  {Mazurek}}, \ and\ \bibinfo {author} {\bibfnamefont {Kevin~J}\ \bibnamefont
  {Resch}},\ }\bibfield  {title} {\enquote {\bibinfo {title} {{Theory of
  high-efficiency sum-frequency generation for single-photon waveform
  conversion}},}\ }\href@noop {} {\bibfield  {journal} {\bibinfo  {journal}
  {Phys. Rev. A}\ }\textbf {\bibinfo {volume} {91}},\ \bibinfo {pages} {033809}
  (\bibinfo {year} {2015})}\BibitemShut {NoStop}%
\bibitem [{\citenamefont {Monmayrant}\ \emph {et~al.}(2010)\citenamefont
  {Monmayrant}, \citenamefont {Weber},\ and\ \citenamefont
  {Chatel}}]{Monmayrant:2010gk}%
  \BibitemOpen
  \bibfield  {author} {\bibinfo {author} {\bibfnamefont {Antoine}\ \bibnamefont
  {Monmayrant}}, \bibinfo {author} {\bibfnamefont {S{\'e}bastien}\ \bibnamefont
  {Weber}}, \ and\ \bibinfo {author} {\bibfnamefont {B{\'e}atrice}\
  \bibnamefont {Chatel}},\ }\bibfield  {title} {\enquote {\bibinfo {title} {{A
  newcomer's guide to ultrashort pulse shaping and characterization}},}\
  }\href@noop {} {\bibfield  {journal} {\bibinfo  {journal} {J. Phys. B: At.
  Mol. Opt. Phys.}\ }\textbf {\bibinfo {volume} {43}},\ \bibinfo {pages}
  {103001} (\bibinfo {year} {2010})}\BibitemShut {NoStop}%
\bibitem [{\citenamefont {de~Riedmatten}\ \emph {et~al.}(2003)\citenamefont
  {de~Riedmatten}, \citenamefont {Marcikic}, \citenamefont {Tittel},
  \citenamefont {Zbinden},\ and\ \citenamefont {Gisin}}]{Riedmatten:2003cl}%
  \BibitemOpen
  \bibfield  {author} {\bibinfo {author} {\bibfnamefont {H}~\bibnamefont
  {de~Riedmatten}}, \bibinfo {author} {\bibfnamefont {I}~\bibnamefont
  {Marcikic}}, \bibinfo {author} {\bibfnamefont {W}~\bibnamefont {Tittel}},
  \bibinfo {author} {\bibfnamefont {H}~\bibnamefont {Zbinden}}, \ and\ \bibinfo
  {author} {\bibfnamefont {N}~\bibnamefont {Gisin}},\ }\bibfield  {title}
  {\enquote {\bibinfo {title} {{Quantum interference with photon pairs created
  in spatially separated sources}},}\ }\href@noop {} {\bibfield  {journal}
  {\bibinfo  {journal} {Phys. Rev. A}\ }\textbf {\bibinfo {volume} {67}},\
  \bibinfo {pages} {022301} (\bibinfo {year} {2003})}\BibitemShut {NoStop}%
\bibitem [{\citenamefont {Kaltenbaek}\ \emph {et~al.}(2006)\citenamefont
  {Kaltenbaek}, \citenamefont {Blauensteiner}, \citenamefont {{\.{Z}}ukowski},
  \citenamefont {Aspelmeyer},\ and\ \citenamefont
  {Zeilinger}}]{Kaltenbaek:2006ch}%
  \BibitemOpen
  \bibfield  {author} {\bibinfo {author} {\bibfnamefont {R}~\bibnamefont
  {Kaltenbaek}}, \bibinfo {author} {\bibfnamefont {B}~\bibnamefont
  {Blauensteiner}}, \bibinfo {author} {\bibfnamefont {M}~\bibnamefont
  {{\.{Z}}ukowski}}, \bibinfo {author} {\bibfnamefont {M}~\bibnamefont
  {Aspelmeyer}}, \ and\ \bibinfo {author} {\bibfnamefont {A}~\bibnamefont
  {Zeilinger}},\ }\bibfield  {title} {\enquote {\bibinfo {title} {{Experimental
  interference of independent photons}},}\ }\href@noop {} {\bibfield  {journal}
  {\bibinfo  {journal} {Phys. Rev. Lett.}\ }\textbf {\bibinfo {volume} {96}},\
  \bibinfo {pages} {240502} (\bibinfo {year} {2006})}\BibitemShut {NoStop}%
\bibitem [{\citenamefont {Eisert}\ \emph {et~al.}(2002)\citenamefont {Eisert},
  \citenamefont {Scheel},\ and\ \citenamefont {Plenio}}]{Eisert:2002wb}%
  \BibitemOpen
  \bibfield  {author} {\bibinfo {author} {\bibfnamefont {J}~\bibnamefont
  {Eisert}}, \bibinfo {author} {\bibfnamefont {S}~\bibnamefont {Scheel}}, \
  and\ \bibinfo {author} {\bibfnamefont {M~B}\ \bibnamefont {Plenio}},\
  }\bibfield  {title} {\enquote {\bibinfo {title} {{Distilling Gaussian states
  with Gaussian operations is impossible}},}\ }\href@noop {} {\bibfield
  {journal} {\bibinfo  {journal} {Phys. Rev. Lett.}\ }\textbf {\bibinfo
  {volume} {89}} (\bibinfo {year} {2002})}\BibitemShut {NoStop}%
\bibitem [{\citenamefont {Bra{\'{n}}czyk}\ \emph {et~al.}(2011)\citenamefont
  {Bra{\'{n}}czyk}, \citenamefont {Fedrizzi}, \citenamefont {Stace},
  \citenamefont {Ralph},\ and\ \citenamefont {White}}]{Branczyk:2011uw}%
  \BibitemOpen
  \bibfield  {author} {\bibinfo {author} {\bibfnamefont {A~M}\ \bibnamefont
  {Bra{\'{n}}czyk}}, \bibinfo {author} {\bibfnamefont {A}~\bibnamefont
  {Fedrizzi}}, \bibinfo {author} {\bibfnamefont {T~M}\ \bibnamefont {Stace}},
  \bibinfo {author} {\bibfnamefont {T~C}\ \bibnamefont {Ralph}}, \ and\
  \bibinfo {author} {\bibfnamefont {A~G}\ \bibnamefont {White}},\ }\bibfield
  {title} {\enquote {\bibinfo {title} {{Engineered optical nonlinearity for a
  quantum light source}},}\ }\href@noop {} {\bibfield  {journal} {\bibinfo
  {journal} {Opt. Express}\ }\textbf {\bibinfo {volume} {19}} (\bibinfo {year}
  {2011})}\BibitemShut {NoStop}%
\bibitem [{\citenamefont {Ben~Dixon}\ \emph {et~al.}(2013)\citenamefont
  {Ben~Dixon}, \citenamefont {Shapiro},\ and\ \citenamefont
  {Wong}}]{BenDixon:2013tk}%
  \BibitemOpen
  \bibfield  {author} {\bibinfo {author} {\bibfnamefont {P}~\bibnamefont
  {Ben~Dixon}}, \bibinfo {author} {\bibfnamefont {J~H}\ \bibnamefont
  {Shapiro}}, \ and\ \bibinfo {author} {\bibfnamefont {FNC}\ \bibnamefont
  {Wong}},\ }\bibfield  {title} {\enquote {\bibinfo {title} {{Spectral
  engineering by Gaussian phase-matching for quantum photonics}},}\ }\href@noop
  {} {\bibfield  {journal} {\bibinfo  {journal} {Opt. Express}\ }\textbf
  {\bibinfo {volume} {21}},\ \bibinfo {pages} {5879--5890} (\bibinfo {year}
  {2013})}\BibitemShut {NoStop}%
\bibitem [{\citenamefont {Dosseva}\ \emph {et~al.}(2014)\citenamefont
  {Dosseva}, \citenamefont {Cincio},\ and\ \citenamefont
  {Branczyk}}]{Dosseva:2014wr}%
  \BibitemOpen
  \bibfield  {author} {\bibinfo {author} {\bibfnamefont {Annamaria}\
  \bibnamefont {Dosseva}}, \bibinfo {author} {\bibfnamefont {Lukasz}\
  \bibnamefont {Cincio}}, \ and\ \bibinfo {author} {\bibfnamefont {Agata~M}\
  \bibnamefont {Branczyk}},\ }\bibfield  {title} {\enquote {\bibinfo {title}
  {{Shaping the spectrum of downconverted photons through optimized custom
  poling}},}\ }\href@noop {} {\bibfield  {journal} {\bibinfo  {journal}
  {arXiv}\ } (\bibinfo {year} {2014})},\ \Eprint
  {http://arxiv.org/abs/1410.7714} {1410.7714} \BibitemShut {NoStop}%
\bibitem [{\citenamefont {Nunn}\ \emph {et~al.}(2013)\citenamefont {Nunn},
  \citenamefont {Wright}, \citenamefont {S{\"o}ller}, \citenamefont {Zhang},
  \citenamefont {Walmsley},\ and\ \citenamefont {Smith}}]{Nunn:2013kf}%
  \BibitemOpen
  \bibfield  {author} {\bibinfo {author} {\bibfnamefont {J}~\bibnamefont
  {Nunn}}, \bibinfo {author} {\bibfnamefont {L~J}\ \bibnamefont {Wright}},
  \bibinfo {author} {\bibfnamefont {C}~\bibnamefont {S{\"o}ller}}, \bibinfo
  {author} {\bibfnamefont {L}~\bibnamefont {Zhang}}, \bibinfo {author}
  {\bibfnamefont {I~A}\ \bibnamefont {Walmsley}}, \ and\ \bibinfo {author}
  {\bibfnamefont {B~J}\ \bibnamefont {Smith}},\ }\bibfield  {title} {\enquote
  {\bibinfo {title} {{Large-alphabet time-frequency entangled quantum key
  distribution by means of time-to-frequency conversion}},}\ }\href@noop {}
  {\bibfield  {journal} {\bibinfo  {journal} {Opt. Express}\ }\textbf {\bibinfo
  {volume} {21}},\ \bibinfo {pages} {15959--15973} (\bibinfo {year}
  {2013})}\BibitemShut {NoStop}%
\bibitem [{\citenamefont {Humphreys}\ \emph {et~al.}(2014)\citenamefont
  {Humphreys}, \citenamefont {Kolthammer}, \citenamefont {Nunn}, \citenamefont
  {Barbieri}, \citenamefont {Datta},\ and\ \citenamefont
  {Walmsley}}]{Humphreys:2014ce}%
  \BibitemOpen
  \bibfield  {author} {\bibinfo {author} {\bibfnamefont {Peter~C}\ \bibnamefont
  {Humphreys}}, \bibinfo {author} {\bibfnamefont {W~Steven}\ \bibnamefont
  {Kolthammer}}, \bibinfo {author} {\bibfnamefont {Joshua}\ \bibnamefont
  {Nunn}}, \bibinfo {author} {\bibfnamefont {Marco}\ \bibnamefont {Barbieri}},
  \bibinfo {author} {\bibfnamefont {Animesh}\ \bibnamefont {Datta}}, \ and\
  \bibinfo {author} {\bibfnamefont {Ian~A}\ \bibnamefont {Walmsley}},\
  }\bibfield  {title} {\enquote {\bibinfo {title} {{Continuous-Variable Quantum
  Computing in Optical Time-Frequency Modes Using Quantum Memories}},}\
  }\href@noop {} {\bibfield  {journal} {\bibinfo  {journal} {Phys. Rev. Lett.}\
  }\textbf {\bibinfo {volume} {113}},\ \bibinfo {pages} {130502} (\bibinfo
  {year} {2014})}\BibitemShut {NoStop}%
\bibitem [{\citenamefont {Reck}\ \emph {et~al.}(1994)\citenamefont {Reck},
  \citenamefont {Zeilinger}, \citenamefont {Bernstein},\ and\ \citenamefont
  {Bertani}}]{Reck:1994dz}%
  \BibitemOpen
  \bibfield  {author} {\bibinfo {author} {\bibfnamefont {Michael}\ \bibnamefont
  {Reck}}, \bibinfo {author} {\bibfnamefont {Anton}\ \bibnamefont {Zeilinger}},
  \bibinfo {author} {\bibfnamefont {Herbert~J}\ \bibnamefont {Bernstein}}, \
  and\ \bibinfo {author} {\bibfnamefont {Philip}\ \bibnamefont {Bertani}},\
  }\bibfield  {title} {\enquote {\bibinfo {title} {{Experimental realization of
  any discrete unitary operator}},}\ }\href@noop {} {\bibfield  {journal}
  {\bibinfo  {journal} {Phys. Rev. Lett.}\ }\textbf {\bibinfo {volume} {73}},\
  \bibinfo {pages} {58--61} (\bibinfo {year} {1994})}\BibitemShut {NoStop}%
\bibitem [{\citenamefont {Gilbert}\ \emph {et~al.}(2006)\citenamefont
  {Gilbert}, \citenamefont {Hamrick},\ and\ \citenamefont
  {Weinstein}}]{Gilbert:2006ba}%
  \BibitemOpen
  \bibfield  {author} {\bibinfo {author} {\bibfnamefont {Gerald}\ \bibnamefont
  {Gilbert}}, \bibinfo {author} {\bibfnamefont {Michael}\ \bibnamefont
  {Hamrick}}, \ and\ \bibinfo {author} {\bibfnamefont {Yaakov~S}\ \bibnamefont
  {Weinstein}},\ }\bibfield  {title} {\enquote {\bibinfo {title} {{Efficient
  construction of photonic quantum-computational clusters}},}\ }\href@noop {}
  {\bibfield  {journal} {\bibinfo  {journal} {Phys. Rev. A}\ }\textbf {\bibinfo
  {volume} {73}},\ \bibinfo {pages} {064303} (\bibinfo {year}
  {2006})}\BibitemShut {NoStop}%
\bibitem [{\citenamefont {Browne}\ and\ \citenamefont
  {Rudolph}(2005)}]{Browne:2005dd}%
  \BibitemOpen
  \bibfield  {author} {\bibinfo {author} {\bibfnamefont {Daniel~E}\
  \bibnamefont {Browne}}\ and\ \bibinfo {author} {\bibfnamefont {Terry}\
  \bibnamefont {Rudolph}},\ }\bibfield  {title} {\enquote {\bibinfo {title}
  {{Resource-Efficient Linear Optical Quantum Computation}},}\ }\href@noop {}
  {\bibfield  {journal} {\bibinfo  {journal} {Phys. Rev. Lett.}\ }\textbf
  {\bibinfo {volume} {95}},\ \bibinfo {pages} {010501} (\bibinfo {year}
  {2005})}\BibitemShut {NoStop}%
\bibitem [{\citenamefont {Knill}\ \emph {et~al.}(2001)\citenamefont {Knill},
  \citenamefont {Laflamme},\ and\ \citenamefont {Milburn}}]{Knill:2001vi}%
  \BibitemOpen
  \bibfield  {author} {\bibinfo {author} {\bibfnamefont {E}~\bibnamefont
  {Knill}}, \bibinfo {author} {\bibfnamefont {R}~\bibnamefont {Laflamme}}, \
  and\ \bibinfo {author} {\bibfnamefont {G~J}\ \bibnamefont {Milburn}},\
  }\bibfield  {title} {\enquote {\bibinfo {title} {{A scheme for efficient
  quantum computation with linear optics}},}\ }\href@noop {} {\bibfield
  {journal} {\bibinfo  {journal} {Nature}\ }\textbf {\bibinfo {volume} {409}},\
  \bibinfo {pages} {46--52} (\bibinfo {year} {2001})}\BibitemShut {NoStop}%
\bibitem [{\citenamefont {Raussendorf}\ and\ \citenamefont
  {Briegel}(2001)}]{Raussendorf:2001js}%
  \BibitemOpen
  \bibfield  {author} {\bibinfo {author} {\bibfnamefont {Robert}\ \bibnamefont
  {Raussendorf}}\ and\ \bibinfo {author} {\bibfnamefont {Hans~J}\ \bibnamefont
  {Briegel}},\ }\bibfield  {title} {\enquote {\bibinfo {title} {{A One-Way
  Quantum Computer}},}\ }\href@noop {} {\bibfield  {journal} {\bibinfo
  {journal} {Phys. Rev. Lett.}\ }\textbf {\bibinfo {volume} {86}},\ \bibinfo
  {pages} {5188--5191} (\bibinfo {year} {2001})}\BibitemShut {NoStop}%
\bibitem [{\citenamefont {Reim}\ \emph {et~al.}(2011)\citenamefont {Reim},
  \citenamefont {Michelberger}, \citenamefont {Lee}, \citenamefont {Nunn},
  \citenamefont {Langford},\ and\ \citenamefont {Walmsley}}]{Reim:2011gr}%
  \BibitemOpen
  \bibfield  {author} {\bibinfo {author} {\bibfnamefont {K.~F.}\ \bibnamefont
  {Reim}}, \bibinfo {author} {\bibfnamefont {P}~\bibnamefont {Michelberger}},
  \bibinfo {author} {\bibfnamefont {K.~C.}\ \bibnamefont {Lee}}, \bibinfo
  {author} {\bibfnamefont {J}~\bibnamefont {Nunn}}, \bibinfo {author}
  {\bibfnamefont {N~K}\ \bibnamefont {Langford}}, \ and\ \bibinfo {author}
  {\bibfnamefont {I~A}\ \bibnamefont {Walmsley}},\ }\bibfield  {title}
  {\enquote {\bibinfo {title} {{Single-Photon-Level Quantum Memory at Room
  Temperature}},}\ }\href@noop {} {\bibfield  {journal} {\bibinfo  {journal}
  {Phys. Rev. Lett.}\ }\textbf {\bibinfo {volume} {107}},\ \bibinfo {pages}
  {053603} (\bibinfo {year} {2011})}\BibitemShut {NoStop}%
\bibitem [{\citenamefont {Nunn}\ \emph {et~al.}(2007)\citenamefont {Nunn},
  \citenamefont {Walmsley}, \citenamefont {Raymer}, \citenamefont {Surmacz},
  \citenamefont {Waldermann}, \citenamefont {Wang},\ and\ \citenamefont
  {Jaksch}}]{Nunn:2007fi}%
  \BibitemOpen
  \bibfield  {author} {\bibinfo {author} {\bibfnamefont {J}~\bibnamefont
  {Nunn}}, \bibinfo {author} {\bibfnamefont {I~A}\ \bibnamefont {Walmsley}},
  \bibinfo {author} {\bibfnamefont {M~G}\ \bibnamefont {Raymer}}, \bibinfo
  {author} {\bibfnamefont {K}~\bibnamefont {Surmacz}}, \bibinfo {author}
  {\bibfnamefont {F~C}\ \bibnamefont {Waldermann}}, \bibinfo {author}
  {\bibfnamefont {Z}~\bibnamefont {Wang}}, \ and\ \bibinfo {author}
  {\bibfnamefont {D}~\bibnamefont {Jaksch}},\ }\bibfield  {title} {\enquote
  {\bibinfo {title} {{Mapping broadband single-photon wave packets into an
  atomic memory}},}\ }\href@noop {} {\bibfield  {journal} {\bibinfo  {journal}
  {Phys. Rev. A}\ }\textbf {\bibinfo {volume} {75}},\ \bibinfo {pages} {011401}
  (\bibinfo {year} {2007})}\BibitemShut {NoStop}%
\bibitem [{\citenamefont {Zheng}\ \emph {et~al.}(2015)\citenamefont {Zheng},
  \citenamefont {Mishina}, \citenamefont {Treps},\ and\ \citenamefont
  {Fabre}}]{Zheng:2015gm}%
  \BibitemOpen
  \bibfield  {author} {\bibinfo {author} {\bibfnamefont {Zhan}\ \bibnamefont
  {Zheng}}, \bibinfo {author} {\bibfnamefont {Oxana}\ \bibnamefont {Mishina}},
  \bibinfo {author} {\bibfnamefont {Nicolas}\ \bibnamefont {Treps}}, \ and\
  \bibinfo {author} {\bibfnamefont {Claude}\ \bibnamefont {Fabre}},\ }\bibfield
   {title} {\enquote {\bibinfo {title} {{Atomic quantum memory for multimode
  frequency combs}},}\ }\href@noop {} {\bibfield  {journal} {\bibinfo
  {journal} {Phys. Rev. A}\ }\textbf {\bibinfo {volume} {91}},\ \bibinfo
  {pages} {031802} (\bibinfo {year} {2015})}\BibitemShut {NoStop}%
\bibitem [{\citenamefont {Alferness}\ \emph {et~al.}(1982)\citenamefont
  {Alferness}, \citenamefont {Ramaswamy}, \citenamefont {Korotky},
  \citenamefont {Divino},\ and\ \citenamefont {Buhl}}]{Alferness:1982vc}%
  \BibitemOpen
  \bibfield  {author} {\bibinfo {author} {\bibfnamefont {R~C}\ \bibnamefont
  {Alferness}}, \bibinfo {author} {\bibfnamefont {V~R}\ \bibnamefont
  {Ramaswamy}}, \bibinfo {author} {\bibfnamefont {S~K}\ \bibnamefont
  {Korotky}}, \bibinfo {author} {\bibfnamefont {M~D}\ \bibnamefont {Divino}}, \
  and\ \bibinfo {author} {\bibfnamefont {L~L}\ \bibnamefont {Buhl}},\
  }\bibfield  {title} {\enquote {\bibinfo {title} {{Efficient Single-Mode Fiber
  to Titanium Diffused Lithium-Niobate Waveguide Coupling for
  Gamma=1.32-Mu-M}},}\ }\href@noop {} {\bibfield  {journal} {\bibinfo
  {journal} {Ieee Journal of Quantum Electronics}\ }\textbf {\bibinfo {volume}
  {18}},\ \bibinfo {pages} {1807--1813} (\bibinfo {year} {1982})}\BibitemShut
  {NoStop}%
\bibitem [{\citenamefont {Luo}\ \emph {et~al.}(2015)\citenamefont {Luo},
  \citenamefont {Herrmann}, \citenamefont {Krapick}, \citenamefont {Brecht},
  \citenamefont {Ricken}, \citenamefont {Quiring}, \citenamefont {Suche},
  \citenamefont {Sohler},\ and\ \citenamefont {Silberhorn}}]{Luo:2015vz}%
  \BibitemOpen
  \bibfield  {author} {\bibinfo {author} {\bibfnamefont {Kai-Hong}\
  \bibnamefont {Luo}}, \bibinfo {author} {\bibfnamefont {Harald}\ \bibnamefont
  {Herrmann}}, \bibinfo {author} {\bibfnamefont {Stephan}\ \bibnamefont
  {Krapick}}, \bibinfo {author} {\bibfnamefont {Benjamin}\ \bibnamefont
  {Brecht}}, \bibinfo {author} {\bibfnamefont {Raimund}\ \bibnamefont
  {Ricken}}, \bibinfo {author} {\bibfnamefont {Viktor}\ \bibnamefont
  {Quiring}}, \bibinfo {author} {\bibfnamefont {Hubertus}\ \bibnamefont
  {Suche}}, \bibinfo {author} {\bibfnamefont {Wolfgang}\ \bibnamefont
  {Sohler}}, \ and\ \bibinfo {author} {\bibfnamefont {Christine}\ \bibnamefont
  {Silberhorn}},\ }\bibfield  {title} {\enquote {\bibinfo {title} {{Direct
  generation of genuine single-longitudinal-mode narrowband photon pairs}},}\
  }\href@noop {} {\bibfield  {journal} {\bibinfo  {journal} {arXiv}\ }
  (\bibinfo {year} {2015})},\ \Eprint {http://arxiv.org/abs/1504.01854}
  {1504.01854} \BibitemShut {NoStop}%
\end{thebibliography}%
\end{document}